\documentclass[12pt]{JHEP3} 

\usepackage{amsmath}
\usepackage{amssymb} 

\usepackage{xspace}
\usepackage{epsfig}
\usepackage{graphics}
\usepackage{subfigure}

\newcommand{\color}[2][x]{}

\usepackage{psfrag}

\newcommand{\ie}{\emph{i.e.}\ }
\newcommand{\eg}{\emph{e.g.}\ }
\newcommand{\cnf}{\emph{cf.}\ }

\def\order#1{{\cal{O}}\left(#1\right)}

\newcommand{\cH}{{\cal H}}

\newcommand{\vProb}{f}
\newcommand{\momConf}{{\cal B}}

\newcommand{\subProc}{\delta}

\newcommand{\jets}{\;\mathrm{jets}}
\newcommand{\jet}{\;\mathrm{jet}}
\newcommand{\GeV}{\;\mathrm{GeV}}

\newcommand{\caesar}{\textsc{caesar}\xspace}

\def\cF{{\cal{F}}}    

\def\cR{{\cal{R}}}
\def\cC{{\cal{C}}}
\def\barcC{{\cal{\bar C}}}
\def\cE{{\cal{E}}}

\def\cR{{\cal{R}}}               

\def\half{\mbox{\small $\frac{1}{2}$}}

\def\CF{C_F}

\def\CA{C_A}

\def\as{\alpha_{{\textsc{s}}}}
\def\gae{{\gamma_{\textsc{e}}}}

\def\cO#1{{\cal{O}}\left(#1\right)}

\def\ee{e^+e^-}

\def\shat{\hat s}

\title{Resummed event shapes at hadron-hadron colliders}
\author{Andrea Banfi\\
  NIKHEF Theory Group, P.O. Box 41882, 1009 DB Amsterdam, 
  The Netherlands.}

\author{Gavin P. Salam\\
  LPTHE, Universities of Paris VI and VII and CNRS UMR 7589, Paris,
  France.}

\author{Giulia Zanderighi\\
  Fermilab, P.O. Box 500, Batavia, IL, US.}

\abstract{ This article introduces definitions for a number of new
  event shapes and jet-rates in hadron-hadron dijet
  production.  They are designed so as
  to be measurable in
  practice at the Tevatron and the LHC, and to be global so that they
  can be resummed with currently available techniques. We explain how
  to vary their sensitivity to beam fragmentation, limiting its impact
  for purely perturbative studies, or deliberately enhancing it so as
  to focus on non-perturbative effects. Explicit next-to-leading
  logarithmic resummed results
  are presented, as obtained with \caesar.  }

\preprint{
  FERMILAB-PUB-04-117-T\\
  LPTHE--04--18\\
  NIKHEF/2004-006\\
  hep-ph/0407287 \\
  July 2004
}

\begin{document}

\section{Introduction}
\label{sec:intro}

Event shapes measure
geometrical properties of the energy flow in QCD
final states. Their conceptual simplicity combined with a remarkable
sensitivity to a range of features of QCD radiation, has led to their
being among the most extensively studied QCD observables, both
theoretically and experimentally. This is especially the case in $\ee$
and DIS collisions, where seminal results include not just a plethora
of measurements of the strong coupling~\cite{Bethke:2002rv}, but also
tests of the colour structure of QCD~\cite{SU3}, detailed validation
of Monte Carlo event
generators~\cite{Herwig,Pythia,Ariadne,TuningDelphi}, 
and profound
insights into the dynamics of hadronisation (for reviews see
\cite{Beneke,DasSalReview}).

Some of these studies consider mean values of event shapes, usually
compared to fixed order perturbation theory, but the most information
is to be gained from event-shape \emph{distributions}. The majority of
data corresponds to events whose energy flow closely resembles that of
the lowest order (Born) event, a region usually associated with a
small value ($V$) of the event shape. This is because large departures
from the Born energy flow are caused by radiation of one or more hard
gluons, each of which costs a power of $\as$. In contrast the dominant
region, $V\ll 1$, is characterised by the presence of multiple soft
and collinear radiation and associated virtual corrections (Sudakov
suppression), which leads to large perturbative terms $(\as\ln^2 V)^n$
that have to be resummed to all orders. Fixed-order and resummed
calculations, both at next-to-leading (NL) accuracy, have been vital
to the full exploitation of the vast amount of data in $\ee$ and DIS.

At hadron colliders, event shapes (and a related class of
observables, jet-res\-olution threshold parameters) have so far received
much less attention, there being for example at the Tevatron only few
experimental measurements \cite{CDF-Broadening,D0Thrust}. Recently
however a number of tools for investigating them beyond leading order
(or leading logarithms) have started to be developed
\cite{JetratesHHResummed,KoutZ0,NLOJET,Nagy03,TriRad,MCFM,BCMN,BSZ03,Caesar}.
This opens up the possibility of using hadron-collider event shapes
for a wide range of quantitative studies.

Compared to $\ee$ (and to some extent DIS) the hadron-collider
experimental environment is more challenging, notably due to the
presence of the underlying event (or equivalently beam fragmentation)
and to the limited detector coverage in rapidity. Theoretically the
situation is also more complex: while in $\ee$ at lowest order there
is radiation from just two outgoing jets, in hadron collider dijet
events for example, at lowest order there is radiation from two
incoming and two outgoing jets, with dependence on incoming parton
distributions as well as a rich structure of interference between
emissions from the different jets
\cite{YuriEtAlMultiJet,BottsSterman,KS,KOS,Oderda,KidonakisOwens}.

The more challenging environment and richer structure, though sources
of technical difficulties, provide many of the motivations for
extending event-shape analyses to hadron colliders. Thus a study of
hadron collider event shapes will not merely be a replication of what
has been done in $\ee$ and DIS (for example with measurements of $\as$)
but will allow novel investigations of both perturbative and
non-perturbative aspects of QCD.

Perturbatively, event-shape distributions are sensitive
to the underlying jet-production channel: an event $p\bar p\to
2\jets$ with an underlying $gg\to gg$ partonic structure will
typically have more radiation than $q\bar q\to q\bar q$, and this will
be reflected in a distribution for that channel that is peaked at
larger values of $V$.

Dijet events will also allow tests of new structures in perturbative
QCD --- 
colour evolution matrices (also known as soft anomalous dimension
matrices), initially developed and studied at Stony Brook 
\cite{BottsSterman,KS,KOS,Oderda,KidonakisOwens}
--- that arise for the first time in events with four or more hard
partons. In processes with three jets (for example), the combined
colour charge of any pair of partons is uniquely determined, via
colour conservation, by the colour charge of the third parton (triplet
for a quark, octet for a gluon). This is important because the pair
colour charge affects the pattern of large-angle soft radiation ---
the fact that it is unique means that one can ignore the evolution of
colour associated with virtual corrections, allowing one to attribute
a `classical' probabilistic interpretation to the resummed virtual
corrections. When there are four or more jets, colour conservation is
insufficient to uniquely fix the combined colour charge of any given
pair of partons, and one needs therefore to account for the `quantum' 
evolution of colour in the QCD amplitudes, and this is accomplished
with the Stony Brook colour evolution matrices. Some first
investigations of their
phenomenological consequences have been given in \cite{Oderda,ApplebySeymour}
in the context of rapidity gaps, however event shapes should allow
more extensive studies, as discussed later in the paper.

On the non-perturbative front it is worth recalling that one of the
most important applications of $\ee$ and DIS event-shape studies in
recent years has been for extracting information about hadronisation,
notably through `power-correction' studies. It has been suggested
\cite{PowerCorrectionPioneers,ShapeFunctions}, using renormalon
\cite{Beneke} inspired techniques, that leading hadronisation effects
essentially act to shift event-shape distributions by an amount of
order $\Lambda_{QCD}/Q$, with $Q$ a hard scale of the process. The
coefficient of this correction is given by the product of two factors:
one of which (denoted $c_V$) is calculable in perturbation theory and
observable dependent, the other one ($\alpha_0$) being fundamentally
non-perturbative, but universal for a range of observables in various
processes. More sophisticated treatments, that attempt to go beyond
leading effects, via `shape functions' have also been developed
\cite{ShapeFunctions}. There have been extensive tests of these ideas
for 2-jet ($q\bar q$) events, with remarkable successes both in $\ee$
and DIS, as reviewed in \cite{DasSalReview}.

Many of the predictions from these renormalon-inspired techniques
coincide with those from a simple Feynman-Field model of
hadronisation~\cite{tube-model}. 
To truly validate them requires that one investigate
also processes with gluon jets and with non-trivial geometries of the
jets (as in multi-jet events). Some studies in $\ee$ and DIS are in
progress, however multi-jet events there are rare (suppressed by
powers of $\alpha_s$). Hadron collider dijet events, in contrast,
provide a natural environment for such tests, because of their
inherent multi-jet structure and the large fraction of gluon jets.

The beam jets do introduce certain complications in these studies,
since their hadronisation (the part that is usually called `underlying
event') is not expected to be fully described by renormalon-inspired
methods. Observables specifically intended for studying beam
fragmentation have been proposed in \cite{MarchWebb88} and examined in
\cite{CDF-Underlying-Event}. Nevertheless, with the aid of a suitably
complementary set of dijet event shapes, it should be possible to
disentangle a number of features of the underlying event, in
particular using approaches as discussed in \cite{KoutZ0}. In this
respect it may also be helpful to consider event shapes in the
somewhat simpler case of Drell-Yan (with and without a jet).

As in $\ee$ and DIS, both fixed-order and resummed perturbative
predictions at NL accuracy are essential in order to fully carry out
these studies.  Usually they are combined together via some form of
`matching' procedure. Fixed-order predictions for arbitrary infrared
collinear safe observables can be obtained with the aid of fixed-order
partonic Monte Carlos \cite{NLOJET,Nagy03,TriRad}. Resummed
predictions, being more intricately 
linked with the details of the observable, had, till now,
traditionally been obtained analytically, by hand. This made it
tedious to consider more than a handful of observables, especially
when dealing with the additional complications of multi-jet topologies
\cite{KoutZ0,eeKout,KoutDIS,AzimDIS,BanfiDasgupta}. Recently however
methods for automating resummations have been developed
\cite{BSZ03,Caesar,BSZ} which enables a much wider range of
studies.

A current technical restriction with these automated resummations (as
well as all fully NLL analytical multijet resummations carried out so
far) is that they apply only to \emph{global} event shapes~\cite{NG1},
\ie observables that are sensitive to radiation in any direction,
notably also close to the beam.\footnote{It is to be noted also that
  the approximations (the angular ordering) contained in current Monte
  Carlo event generators such as Herwig \cite{Herwig} or Pythia
  \cite{Pythia} are at their most accurate for global observables
  (Ariadne \cite{Ariadne} in contrast has similar accuracy for global
  and non-global observables) --- this means that global observables should
  be particularly suited also for the precise validation and tuning of
  event generators at hadron colliders.} The only event shape
distribution that has so far been measured at the Tevatron, a
transverse thrust \cite{D0Thrust}, does not satisfy this property.

One of the main purposes of this article is therefore to introduce and
resum a set of global event shapes and jet-threshold resolution
parameters.  These include observables defined directly
in terms of all particles in the event. Such definitions might seem
particularly unsuited to practical uses, given that experimental
detectors have limited angular reach close to the beam. We shall,
however, discuss the impact on comparisons to theory of the limited
experimentally available rapidity range, arguing that this is not an
unsurmountable obstacle.

We shall also introduce observables defined just in terms of particles
in the central region of the detector, but for which one arranges an
\emph{indirect} sensitivity to the effects of all other remaining
particles. Though these definitions eliminate the problems associated
with limited detector reach, one should be aware that they are subject
to certain extra 
theoretical (and potentially also experimental) problems.

Throughout our discussions we shall refer to resummed results as
obtained with the Computer Automated Expert Semi-Analytical Resummer
(\caesar)~\cite{Caesar}. We shall also consider observable-specific
issues related to the sensitivity to the underlying event.

\section{General considerations}
\label{sec:general}

In order to provide the background for the discussion of the various
observables proposed later in this paper, it is helpful to
first introduce some general concepts and notation.

We shall consider here event shapes, which for any number $N$ of
final-state particles, are defined in terms of some function
$V(q_1,\ldots,q_N)$ of the final-state four-momenta $q_1,\ldots,q_N$.
These event shapes will provide a continuous measure of the extent to
which a general event's energy-momentum flow differs from that of a
lowest order (Born) event.

Typically one wishes to consider the event shape only for events that
are sufficiently hard, requiring for example jets above some minimum
transverse energy threshold $E_{t,\min}$. We will denote this kind of
hardness selection cut by a function $\cH(q_1,\ldots,q_N)$, equal to
$1$ for events that pass the cuts and $0$ otherwise. One can then
define the cross section for events that pass the cuts,
\begin{equation}
  \label{eq:sigmacut}
  \sigma_\cH = \sum_{N} \int d\Phi_N\,
  \frac{d\sigma_N}{d\Phi_N}\,
  \cH(q_1,\ldots,q_N)\,,
\end{equation}
where $d\sigma_N/d\Phi_N$ is the differential cross section for
producing $N$ particles in some configuration $\Phi_N$. While
individual $N$-particle cross sections are, of course,
infrared-collinear (IRC) unsafe, the combined summed and integrated
cross section is safe for any IRC safe function $\cH$.

One also defines the partial integrated cross section $\Sigma_\cH(v)$ for
events that pass the cut and for which additionally the event shape
observable is smaller than some value $v$,
\begin{equation}
  \label{eq:SigmaIntcut}
  \Sigma_\cH(v) = \sum_{N} \int d\Phi_N
  \frac{d\sigma_N}{d\Phi_N} \,\Theta(v - V(q_1,\ldots,q_N))\,
  \cH(q_1,\ldots,q_N)\,.
\end{equation}
The differential normalised distribution for the event shape is then
given by
\begin{equation}
  \label{eq:diffdist}
  \frac{1}{\sigma_\cH} \frac{d\Sigma_\cH(v)}{dv}\,.
\end{equation}

\subsection{Resummation}
\label{sec:resummation}

Resummations are relevant in the region of small $v$, where
logarithmically enhanced contributions, $(\as \ln ^2 v)^n$, are large
at all orders, making fixed-order predictions unreliable. The fact
that the event shape is small implicitly means that the events
resemble the Born event. This allows one to write a factorisation
formula in which the resummed prediction for a given scattering
channel $\subProc$ (for example $qq\to qq$, $qq\to gg$,\ldots)
$\Sigma_{\cH,\subProc}(v)$ is expressed as the integral, over Born
momentum configurations $\momConf$, of the product of the differential
Born cross section $d\sigma_{\subProc}/d\momConf$ and the resummed
probability $f_{\momConf,\subProc}(v)$ that the event shape has a
value smaller than $v$ for that given Born configuration,
\begin{equation}
  \label{eq:Sigmacut_resummed}
  \Sigma_\cH(v) = \sum_{\subProc}
  \Sigma_{\cH,\subProc}(v)\,,
  \qquad \Sigma_{\cH,\subProc}(v) =\int d\momConf\,
  \frac{d\sigma_\subProc}{d\momConf} \,\vProb_{\momConf,\subProc}(v)\,
  \cH(p_{3},p_{4})\,,
\end{equation}
where we use the convention that $p_3$ and $p_4$ denote the momenta of
the outgoing Born partons (while $p_1, p_2$ are the incoming hard
partons). 

The resummed probability $\vProb_{\momConf,\subProc}(v)$ can often be
written in an exponential form \cite{CSS,CTTW}
\begin{equation}
  \label{eq:vProb-general}
  \vProb_{\momConf,\subProc}(v) = \exp\left[ L g_1(\as L) + g_2(\as L)
  + \as g_3(\as L) + \cdots\right]\,, \qquad\quad L = \ln \frac1v\,,
\end{equation}
where $L g_1(\as L)$ resums leading-logarithmic (LL) terms, $ \as^n
L^{n+1}$, $g_2(\as L)$ resums next-to-leading logarithmic (NLL) terms,
$\as^n L^n$, and so on. Current state-of-the-art is NLL resummation.
The exact form of the LL and NLL functions depends on the observable
under consideration. 

One can characterise a given observable by its functional dependence
on the momentum of a \emph{single} soft emission, collinear to one of
the hard (`Born') partons in the event. For all known event shapes,
this can be written as
\begin{equation}
  \label{eq:parametricform}
  V(\{{\tilde p}\}, k)=
  d_{\ell}\left(\frac{k_t^{(\ell)}}{Q}\right)^{a_\ell}
  e^{-b_\ell\eta^{(\ell)}}\, 
  g_\ell(\phi)\>,
\end{equation}
where $\{{\tilde p}\}$ denotes the Born momenta (including recoil
effects) and $k$ is the soft collinear emission; $k_t^{(\ell)}$ and
$\eta^{(\ell)}$ denote respectively its transverse momentum and
rapidity, as measured with respect to the Born parton (`leg') labelled
$\ell$; $\phi$ is the azimuthal angle of the emission with respect to
 a suitably defined event plane (when relevant);
 and $Q$ is the hard scale of the problem.  For
hadron-collider dijet events we will use the convention $\ell=1,2$ for
the incoming hard partons and $\ell=3,4$ for the outgoing hard
partons.

The values of the coefficients $a_\ell$, $b_\ell$ and the form of the
function $g_\ell(\phi)$ in eq.~(\ref{eq:parametricform}) are among the
main characteristics of the observable that enter in the resummation
formula eq.~(\ref{eq:vProb-general}). For example the leading
logarithmic function $Lg_1(\as L)$ depends only on the $a_\ell$ and
$b_\ell$ values, as can be illustrated from its expansion
\begin{equation}
  \label{eq:Lg1-expansion}
  Lg_1(\as L) = -\sum_\ell \frac{C_\ell}{a_\ell(a_\ell +
    b_\ell)}\frac{\as L^2}{\pi}  +
  \order{\as^2 L^3}\,,
\end{equation}
where $C_\ell$ is the colour charge ($\CF$ or $\CA$) of hard parton
$\ell$. For the observables we consider here --- of the continuously
global \cite{NG1,DiscontGlobal} variety --- the $a_\ell$ are all
equal, $a_1 = a_2 = \ldots \equiv a$ (and the $d_\ell$ are all
non-zero).  Because of the appearance of the hard-parton colour
factors in eq.~(\ref{eq:Lg1-expansion}), the LL terms depend also on
the underlying hard-scattering channel $\subProc$ (\eg $qq\to qq$ as
opposed to $gg\to gg$), though not on the particular momentum
configuration $\momConf$ of the hard partons.

The NLL terms are somewhat more complex. One can often separate them
into two pieces,
\begin{equation}
  \label{eq:g2-separated}
  g_2(\as L) = g_{2s}(\as L) + \ln \cF(R'(\as L))\,.
\end{equation}
The first term $g_{2s}(\as L)$ depends just on the `single-emission'
parameters of eq.~(\ref{eq:parametricform}) (including $d_\ell$ and
$g_\ell(\phi)$), and on the Born momenta $\momConf$ and the channel
$\subProc$. It contains, among other contributions, the dependence on
the Stony Brook colour evolution matrices that were mentioned in the
introduction.

The second term in eq.~(\ref{eq:g2-separated}), $\cF(R')$, is the only
part of the NLL resummed formula that is affected by the observable's
dependence on multiple emissions. It is a function of
\begin{equation}
  \label{eq:Rprime}
  R'(\as L) \equiv -\,\partial_L L g_1(\as L)\,.
\end{equation}
Schematically, for $R'\ll 1$ ($L$ not too large, $v$ moderately small)
there is one soft and collinear emission in the event that is harder
than all other emissions. This hardest emission gives the dominant
contribution to the value of the observable\footnote{Strictly speaking
  this is only true for \emph{recursively} infrared-collinear safe
  observables
  \cite{BSZ03,Caesar} --- most event shapes are in this class. } 
and only the single-emission properties of the observable are
relevant, so $\cF(R')\simeq 1$. For $R'\sim 1$ ($L \sim 1/\as$, $v\ll
1$) there are typically $R'+1$ similarly hard gluons that dominate
the structure of the event. If multiple emissions tend to increase the
value of the event shape, then for a given value of the event shape,
this must be compensated for by an extra suppression of emissions, \ie
$\cF(R') < 1$. Conversely if the effects of multiple emissions tend to
cancel out, this is compensated by reducing the overall amount of
suppression, $\cF(R') > 1$. The function $\cF(R')$ often depends on
the underlying scattering channel $\subProc$, but not on the hard
momentum configuration.

Throughout this paper, much of the discussion of the properties of
observables will be framed in terms of the values of the $a_\ell$,
$b_\ell$, $d_\ell$ and the forms of $g_\ell(\phi)$ and $\cF(R')$.

\subsection{Experimental issues}
\label{sec:experimental}

Our understanding of the current Tevatron detectors
\cite{CDF-RunII,D0-RunIIb,PrivJoey,PrivD0} is that typically they have
good measurement capabilities in a central detector region, up to
about $2.5$--$3.0$ units of rapidity $\eta$, as well as a reasonable
degree of measurement capability in the semi-forward region up to a
rapidity of about $3.5$, with much more limited capabilities up to a
rapidity of about $4.5$--$5$. The LHC detectors \cite{ATLASTDR,CMSTDR}
should provide good coverage up to about $5$ units of rapidity.

A limit on the maximum reach in rapidity can be a source of problems
because the most natural way of defining global observables is
directly in terms of all particle momenta in the event. Introducing a
cut on the maximum rapidity at which one measures particles
potentially creates a mismatch between the `ideal' theoretical
definition and the measurement.

It was pointed out however, in \cite{KoutZ0}, that, for sufficiently large
values of the maximum accessible rapidity, $\eta_{\max}$, the excluded
kinematic region gives at most a small contribution to the observable,
which for the discussion of a general observable,
eq.~(\ref{eq:parametricform}), corresponds to an amount of order
$e^{-(a+b_{\min})\eta_{\max}}$ where $b_{\min} = \min\{b_1, b_2\}$.
Accordingly, as long as one is in a region where the observable's
value is larger than this, equivalently
\begin{equation}
  \label{eq:lnV-reach}
  L \lesssim (a +
  b_{\min})\eta_{\max}\,,
\end{equation}
the resummed prediction should remain valid.  One of the features of
the resummed results that we shall therefore examine closely in the
subsequent sections, is the typical values taken by the observable.  As
long as the cross section outside the region eq.~(\ref{eq:lnV-reach})
is small, the effect of the limited experimental rapidity reach should
be negligible. As a reference value for the maximum attainable
rapidity, to be used in eq.~(\ref{eq:lnV-reach}) when comparing with
actual resummed results, we shall take $\eta_{\max} = 3.5$.

A point worth mentioning is that measurement of the contribution
to an event shape from the forward region may be somewhat simpler than
detailed jet studies in that region. This is because many event shapes'
sensitivity to the forward region is simply in terms of the total
transverse energy deposited there, so that it is not necessary to
resolve the detailed $\eta$--$\phi$ structure of the energy
deposition.

An alternative to defining global observables in terms of all momenta
in the event, to be discussed in section~\ref{sec:ind-glob}, is to use
particles only in some restricted central region $\cC$ while
introducing an indirect sensitivity to momenta outside that region
through the addition of the total vector sum of transverse energy in
$\cC$, or equivalently, the missing transverse energy. In such cases,
the critical experimental issue will no longer be the rapidity reach
of the detector, but rather the accuracy with which the missing
transverse energy can be determined. If the missing transverse energy
can be determined to within an error $\delta E_\perp$, then the accessible
region for studying the event shape will be roughly $L \lesssim a \ln
(E_{\perp,jet}/\delta E_\perp)$, where $E_{\perp,jet}$ is the jet
transverse energy.

\subsection{Event selection cuts}
\label{sec:event-select}

The selection cuts that we propose for event-shape studies are as
follows. Using some IRC safe jet-algorithm, one should first select
events where the two jets with highest transverse energy ($E_{\perp,1}>
E_{\perp,2} > \ldots$) are both central, $|\eta_{\mathrm{jets}\,1,2}|
< \eta_\mathrm{c}$, where the limit of the central region
$\eta_\mathrm{c}$ is taken to be of the order of about $0.5$ to $1$.

Restricting the jets to the central region is advantageous
theoretically, because it eliminates large subleading corrections that
are likely to be associated with forward jets. It should also be
advantageous experimentally, insofar as the measurement of the event
shape needs particularly fine resolution in rapidity and azimuth in
the region containing the jets.

A cut should also be placed on the hardness of the jets. To avoid
problems of perturbative convergence of $\sigma_\cH$ that arise with
the use of
symmetric $E_\perp$ cuts on the jets~\cite{KK,symm-cuts}, one can for
example place a cut on the transverse energy $E_{\perp,1}$ of the
hardest jet, $E_{\perp,\min} < E_{\perp,1} < E_{\perp,\max}$.

At the (NLL) accuracy that we discuss in this paper, the details of
the jet algorithm used to identify the two hardest jets, be it of cone
\cite{RunII-jet-physics} or cluster
\cite{JetratesHH-CDSW,JetratesHH-EllisSoper} type, do not modify the 
predictions. This is a consequence of the continuous globalness of the
observables and the fact that we consider just the soft and collinear
limit for the observable. At higher logarithmic accuracy, there are
reasons to believe that the partial integrated cross section
$\Sigma_\cH(v)$ remains independent of the jet algorithm. However
since the total selected cross section, $\sigma_\cH$, does depend on
the jet algorithm at NLO, so does the ratio eq.~(\ref{eq:diffdist}) at
NNLL. In the region of finite $v$, $\Sigma_\cH(v)$ itself also depends
on the jet algorithm.

In the plots shown throughout this paper, we take $\eta_c = 0.7$,
$E_{\perp,\min} = 50\GeV$ and place no limit of $E_{\perp,\max}$. We
assume a Tevatron run II regime, $p\bar p$-collisions at the
centre-of-mass energy $\sqrt{s} = 1.96$TeV and use the CTEQ6M parton
density set~\cite{CTEQ}, corresponding to $\alpha_s(M_Z) = 0.118$.

We set the factorisation and renormalisation scales to be the sum of
the transverse energies of the two most energetic jets. Future work
\cite{BSZMatching} will include fixed-order matching and will explore a
range of alternative scales and perform a systematic study of the
dependence on these scales.

\section{Directly global observables}
\label{sec:dir-glob}

We will first consider observables that are defined in terms of
\emph{all} emissions in the event. This direct sensitivity to all
emission momenta is the origin of the name `directly' global. There
will be two groups of observables --- those that can naturally be
defined in terms of all emissions in the event, and those which would
more naturally be defined in terms of emissions in some central
region, to which we add an extra term that is sensitive to forward
emissions.  In this section we will consider the first class of
observables, while the second kind will be discussed in
section~\ref{sec:rho-brd-etc}.

\subsection{Transverse thrust (a detailed example)}
\label{sec:tau-perp}

We give here a detailed discussion of the 'transverse thrust' event
shape. Since the issues that arise are quite similar for nearly all
event shapes, the discussion for the other event shapes that we define
afterwards will be somewhat briefer, restricted to highlighting
important differences compared to the transverse thrust.

The most obvious extension of the various $\ee$ event shapes is to
define a thrust $T_{\perp,g}$ in the transverse plane,
\begin{equation}
  \label{eq:Ttg}
  T_{\perp,g} \equiv \max_{\vec n_T} \frac{\sum_i |{\vec
      q}_{\perp i}\cdot {\vec
      n_T}|}{\sum_i q_{\perp i}}\,,
\end{equation}
where the sum runs over \emph{all} particles $q_i$ in the final state,
$\vec q_{\perp i}$ represents the two momentum components transverse
to the beam\footnote{We distinguish between `$\perp$' which always
  refers to a transverse momentum with respect to the beam direction,
  and `$t$', used in eq.~(\ref{eq:parametricform}), which is a
  transverse momentum with respect to a given leg.} and $\vec n_T$ is
the transverse vector that maximises the projection.

Variants of the transverse thrust based just on particles in a
restricted central region have been measured in \cite{D0Thrust} and
calculated at fixed order in \cite{Nagy03}. Those variants are
discontinuously global ($a_{1,2}\ne a_{3,4}$), and so beyond the scope
of the current 
automated resummation technology.

\TABLE{
 \begin{tabular}{| c | c | c | c | c | c |}
 \hline
 leg $\ell$ & $a_{\ell}$ & $b_{\ell}$ & $g_{\ell}(\phi)$ & $d_{\ell}$ & $ \langle
 \ln g_{\ell}(\phi) \rangle$ \\
 \hline
 \hline
1 &  1 & $  0$ & $1-|\cos\phi|^*$ &  $1/\sin\theta^*$ & $  -4G/\pi-\ln 2^*$ \\
 \hline
2 &  1 & $  0$ & $1-|\cos\phi|^*$ &  $1/\sin\theta^*$ & $   -4G/\pi-\ln 2^*$ \\
 \hline
3 &  1 & $  1$ &$\sin^2\phi$ &    $1/\sin^2\theta^*$ & $-2\ln 2 $ \\
 \hline
4 &  1 & $  1$ &$\sin^2\phi$ &    $1/\sin^2\theta^*$ & $-2\ln 2 $ \\
 \hline
 \end{tabular}
  \caption{Leg properties for $\tau_{\perp,g}$. Here $G \simeq 0.915965594$ is
    Catalan's constant. Starred entries (here and in subsequent
    tables) indicate quantities that have been determined only
    numerically by \caesar, and for which we have manually provided
    the full analytical information.}
  \label{tab:tautg}
}
\FIGURE{
\psfrag{L = 1,2}{\small $\ell=1,2$}
\psfrag{L = 3,4}{\small $\ell=3,4$}
\psfrag{L}{$\ell$}
  \includegraphics[width=0.45\textwidth]{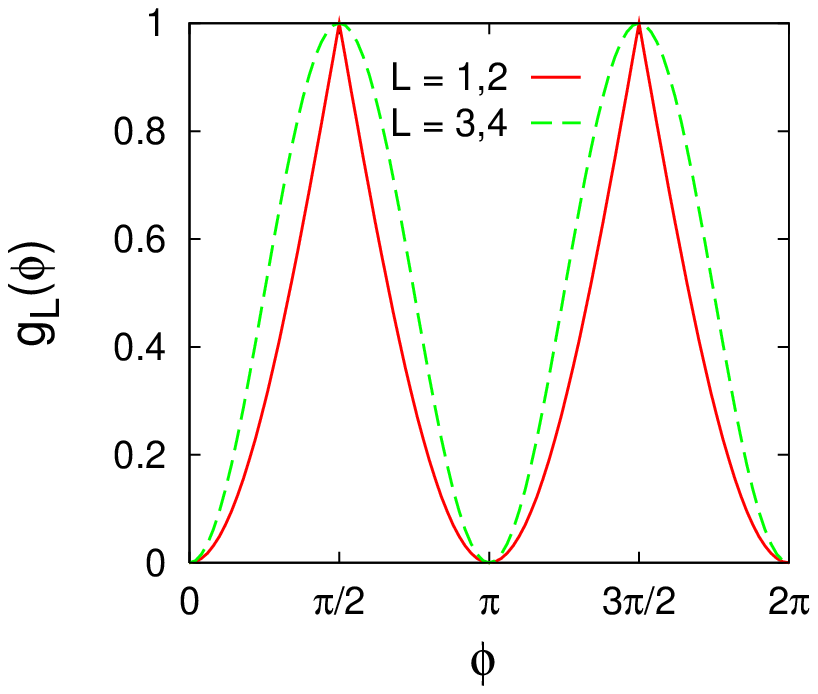}
  \caption{Dependence $g_\ell(\phi)$ of $\tau_{\perp,g}$ on the azimuthal
    angle $\phi$ of a single emission collinear to each leg $\ell$.}
  \label{fig:tautg-phi}
} 

The transverse thrust as defined in eq.~(\ref{eq:Ttg}) has already been
used as brief example of the capabilities of \caesar in \cite{BSZ03},
but here we discuss it more extensively. In the $2+2\jet$ limit it
tends to $T_{\perp,g}=1$. Accordingly the observable that one resums
is $\tau_{\perp,g} \equiv 1 - T_{\perp,g}$.  Its dependence on soft
and collinear emissions is given by eq.~(\ref{eq:parametricform}) with
the parameters shown in table~\ref{tab:tautg}.  The fact that
$b_\ell=0$ for the incoming legs $\ell=1,2$ means that along the beam
direction the observable is sensitive to the emitted transverse
momentum uniformly at all rapidities. In contrast for the outgoing
legs $\ell=3,4$, we have $b_\ell=1$, meaning that the effect of emissions
close to the legs is suppressed. From eq.~(\ref{eq:Lg1-expansion}), one
sees that this translates to smaller LL contributions from the
outgoing jets than for the incoming jets.

The observable's dependence on the azimuthal angle $\phi$ of a single
emission, $g_\ell(\phi)$, is shown in fig.~\ref{fig:tautg-phi}. For
both incoming and outgoing legs, the observable is most sensitive to
radiation perpendicular to the event plane,
$\phi=\frac{\pi}{2},\frac{3\pi}{2}$. The analytical form for
$g_\ell(\phi)$ has been established by \caesar only for the outgoing
legs, it being given numerically for the incoming legs (though the
analytical form is straightforward to derive by hand, and reads
$g_{1,2}(\phi)=1-|\cos\phi|$).

In order to get an understanding of the coefficients $d_\ell$, one
needs to know the value of the hard scale $Q$ in
\eqref{eq:parametricform}. For the present analysis (and for all other
analyses in this paper) we have chosen to set this scale to the
partonic centre of mass energy of the hard collision$ \sqrt{\hat s}$
and have considered a hard configuration in which the scattering angle
$\theta$ between the outgoing jets and the beam direction (in the
hard-scattering centre of mass) satisfies $\cos \theta=0.2$.  \caesar
would provide in table~\ref{tab:tautg} the numerical value
of the coefficient $d_\ell$ corresponding to this reference hard
configuration. 
However,  for all the observables considered in this paper,
we report the functional dependence of $d_\ell$ 
on the angle $\theta$,
as well as the explicit form of $g_\ell(\phi)$, since they can be 
derived from simple analytical considerations (or by examining
the numerical results provided by \caesar). 
To distinguish then the genuine output of \caesar from the
quantities obtained by hand, the latter are highlighted with an asterisk.

Table~\ref{tab:tautg} also includes the result for $\langle\ln
g_\ell(\phi)\rangle = \int \frac{d\phi}{2\pi} \ln g_\ell(\phi)$. This
is of interest insofar as it is actually the combination $\ln \bar
d_\ell\equiv \ln 
d_\ell + \langle\ln g_\ell(\phi)\rangle$ that appears in the
resummation formulae, rather than $d_\ell$ or $g_\ell(\phi)$
separately.

In order to complete the information needed for the resummation,
\caesar evaluates also the function $\cF(R')$ that appears in
eq.~(\ref{eq:g2-separated}). The transverse thrust (strictly,
$\tau_{\perp,g}$) has the property of additivity, meaning that in the
presence of many emissions, the value of the observable is simply the
sum of the values that the observable would take for each emission
individually. For observables with this property, $\cF(R')$ is known
analytically, 
\begin{equation}
  \label{eq:additive-cF}
  \cF(R')= \frac{e^{-\gae R'}}{\Gamma(1 + R')}\,.
\end{equation}
This completes the information needed so as to make resummed
predictions for the transverse thrust.

\DOUBLEFIGURE{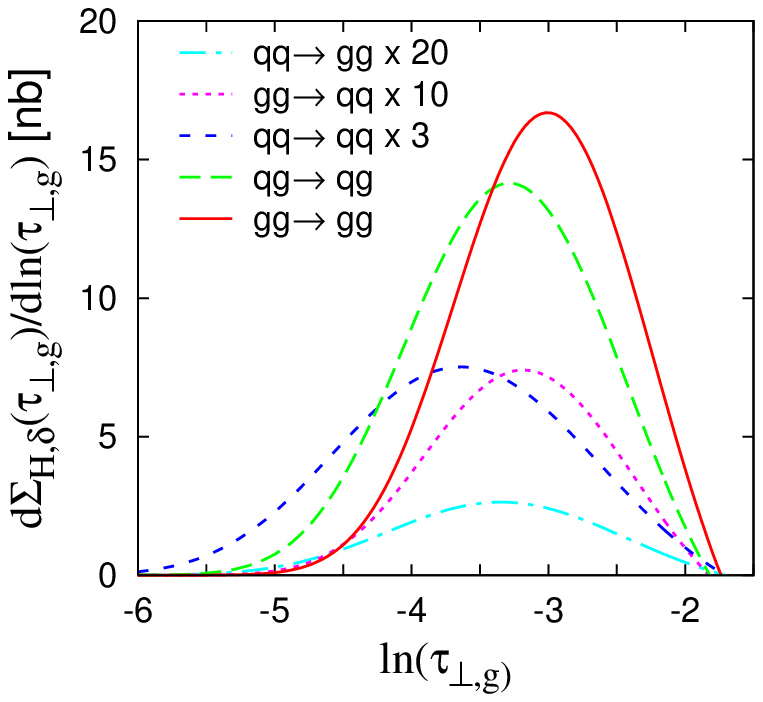,width=0.48\textwidth}%
{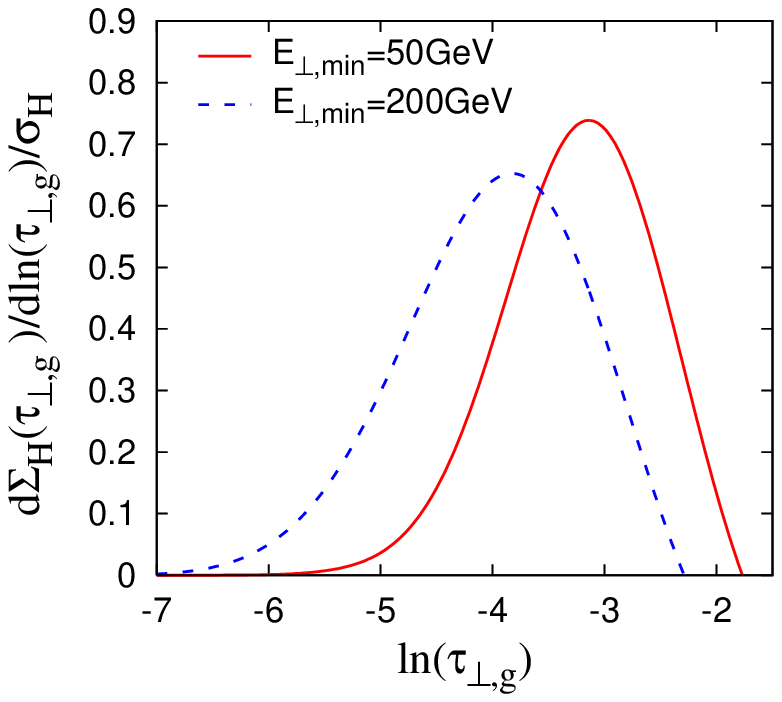,width=0.48\textwidth}%
{The differential cross section $d\Sigma_\cH/d\ln \tau_{\perp,g}$
  separated into the different hard-scattering channels.
  \label{fig:tautg-dists-channels}
}%
{The differential cross section for $\tau_{\perp,g}$, summed over
  channels summed over channels, for two $E_{\perp,\min}$ cuts.
  \label{fig:tautg-dists-sum}}

The resulting differential cross sections for different partonic
scattering channels are shown in
figure~\ref{fig:tautg-dists-channels}, where $q$ denotes a generic
fermion (quark or antiquark) and $qg \to qg$ indicates any process
with incoming fermion and gluon.  They have been obtained by
integrating over all hard events that satisfy the cuts discussed in
section~\ref{sec:event-select}. The $d_\ell$ are redetermined for each
hard configuration, since they depend on the kinematics of the event.
While in table~\ref{tab:tautg} the arbitrary reference hard scale was
taken to be $\sqrt{\hat s}$, hereon, for the purposes of calculating
differential cross sections we choose $Q$, as well as the
renormalisation and factorisation scales, to be the sum of the two
hardest jet transverse energies, which is closer to the virtuality of
the exchanged parton in the hard scattering. Since the distributions
are purely resummed, without the inclusion of the exact fixed-order
contribution (matching, left to future work \cite{BSZMatching}), they
are reliable only at small values of $\tau_{\perp,g}$. For
$\tau_{\perp,g}\sim1$, the differential cross sections are negative,
this being a reflection of a breakdown of the resummation approach in
that region.

At small $\tau_{\perp,g}$ the differential cross sections of
fig.~\ref{fig:tautg-dists-channels} have somewhat different shapes for
the various hard-scattering channels ($\subProc$). This is a
consequence of the different colour factors multiplying the LL terms
in eq.~(\ref{eq:Lg1-expansion}), the LL contribution being larger for
channels with more hard gluons --- such channels tend to radiate more,
so the distribution is peaked at larger values of $\tau_{\perp,g}$.

Combining the different hard-scattering channels, one obtains the full
resummed distribution, fig.~\ref{fig:tautg-dists-sum}. In
section~\ref{sec:experimental} we discussed the consequences of
limited experimental rapidity coverage for comparisons between (fully
global) theory and (partially global) data. The main result was that
the resummation remains valid, in the range limited by
eq.~(\ref{eq:lnV-reach}), which here translates to $L \lesssim
\eta_{\max}$. A subtlety arises in the translation between
$\tau_{\perp,g}$ and the value of $L$: as discussed in
appendix~\ref{sec:scale-choices}, in eq.~(\ref{eq:vProb-general}) we
resum not $\ln 1/\tau_{\perp,g}$ but actually $\ln
1/(X\tau_{\perp,g})$, where $X$ is chosen so as to cancel the average
value of $d_\ell g_\ell(\phi)$. For the hard events satisfying the
cuts in section~\ref{sec:event-select} this corresponds to $X \simeq
5$. This is a convention, adopted from $\ee$ \cite{cpar_res} and DIS
\cite{DiscontGlobal}, intended to minimise spurious subleading
logarithms associated with the potentially arbitrary normalisation of
one's observable.\footnote{Readers familiar also with \cite{BSZ03} may
  have noticed that the cross sections shown in fig.~1 (there) and
  fig.~\ref{fig:tautg-dists-channels} (here) differ considerably in
  normalisation, even though they apply to the same observable. This
  is because in \cite{BSZ03} we used different cuts ($\eta_c = 1.0$
  rather than $0.7$), but also because there we used $X=1$ and
  $Q=\sqrt{\shat}$, both choices being associated, for this
  observable, with large subleading logarithms.}
Since for the discussion of section~\ref{sec:experimental}, it is
$L\equiv \ln 1/(X\tau_{\perp,g})$ that determines the maximum value of
rapidity that is relevant in the resummation, eq.~(\ref{eq:lnV-reach})
should be rewritten
\begin{equation}
  \label{eq:lnV-reach-with-X}
  \ln v \gtrsim -(a + b_\ell) (\eta_{\max} + \langle \ln X
  \rangle)\,.
\end{equation}
Assuming $\eta_{\max}\simeq 3.5$, one obtains the limit $\ln\tau_{\perp,g} \gtrsim -5$, suggesting that the distribution is nearly
entirely in a region where the limited experimental rapidity reach
should be unimportant. If however one goes to higher energies, such as
$E_{\perp,1} > 200\GeV$ as shown in fig.~\ref{fig:tautg-dists-sum},
channels with smaller colour factors start to dominate (\eg $qq
\to qq$), because one samples the parton distributions at higher
$x$; this together with the smaller value of $\as$ leads to the
$\tau_{\perp,g}$ distribution being dominated by lower values of the
observable.

A final point regarding $\tau_{\perp,g}$ concerns its sensitivity to
the underlying event. To parametrise this sensitivity it is useful, as
was done in \cite{KoutZ0}, to take a simple model \cite{MarchWebb88}
in which particles from the underlying event have a spectrum
$dn^{(u.e.)}/dk_\perp d\eta$ that is independent of $\eta$ and of the
underlying hard scattering process. Since $\tau_{\perp,g}$ is
additive, and sensitive along the incoming legs to $k_\perp
g_{1,2}(\phi)$ (see tab.~\ref{tab:tautg}), it receives a mean
contribution from the underlying event
\begin{subequations}
  \label{eq:tautg-underlying-event}
\begin{align}
  \label{eq:tautg-underlying-event-int}
  \langle \delta^{(u.e.)}  \tau_{\perp,g}\rangle &= \left(
    \int_{-\eta_{\max}}^{-\eta_{\max}} d\eta + \order{1} \right)
  \int dk_\perp \frac{d\phi}{2\pi} \frac{dn^{(u.e.)}}{d\eta dk_\perp}
  \frac{k_\perp g_\ell(\phi)}{E_{\perp,1}+E_{\perp,2}}
  \\
  &= \frac{\langle k_\perp^{(u.e.)}\rangle \langle
    g_{1,2}(\phi)\rangle}{E_{\perp,1}+E_{\perp,2}} \left(2\eta_{\max}
    + \order{1}\right)\,,
\end{align}
\end{subequations}
where $\langle k_\perp^{(u.e.)}\rangle$ is the mean transverse
momentum per unit rapidity coming from the underlying event. The
unspecified contribution of $\order{1}$ reflects the fact that for
this qualitative discussion, we have not taken into account the
details of the observable's sensitivity to emissions close to the
outgoing jets. Though the result is for the mean effect of the
underlying event, expectations based on other studies of
non-perturbative effects \cite{DokWebDists} suggest that the
distribution of $\tau_{\perp,g}$ will simply be shifted by $\langle
\delta^{(u.e.)}  \tau_{\perp,g}\rangle$. Because of the
proportionality to $\eta_{\max}$, the effect will be large, making
$\tau_{\perp,g}$ a good observable for testing models of the
underlying event. For example the effects of the underlying event
could depend on the underlying hard scattering channel,\footnote{One of
  us (GPS) wishes to thank B.~R.~Webber for discussions on this
  point.} and it might then be possible, from the event-shape
distribution, to establish whether or not this is the case.

\subsection{Thrust minor}
\label{sec:tmin}

Given the transverse thrust axis $\vec n_T$, one can define a directly
global thrust minor,\footnote{In the literature
  \cite{CDF-Broadening,Nagy03}, such an observable, with an additional
  cut on the rapidities of measured particles, \cnf
  eq.~(\ref{eq:Tm-CE}), has been referred to as a broadening. Our
  choice of nomenclature is in analogy with the thrust minor as
  defined in $\ee$ \cite{MarkJPhysRep}. }
\begin{equation}
  \label{eq:Tmg}
  T_{m,g} \equiv \frac{\sum_i |q_{xi}|}{\sum_i q_{\perp i}}\,,
\end{equation}
where the $x$ direction is defined as that perpendicular to the beam
and to the global transverse thrust axis, which together define the
event plane. The thrust minor can be viewed as a measure of the
out-of-plane momentum, and is quite similar to the observable resummed
analytically in \cite{KoutZ0} for Drell-Yan plus jet production.

One can also normalise the thrust minor to $E_{\perp,1}+E_{\perp,2}$.
This will not modify the resummation, but it will affect
the distribution at values of $T_{m,g}$ of order $1$. (Such a change
cannot be made for the $T_{\perp,g}$, because it would destroy the
positive definiteness of $\tau_{\perp,g} = 1-T_{\perp,g}$).

\TABLE{
\begin{tabular}{| c | c | c | c | c |}
 \hline
 leg $\ell$ & $a_{\ell}$ & $b_{\ell}$ & $g_{\ell}(\phi)$ & $d_{\ell}$
\\
 \hline
 \hline
1 &  1 & $  0$ & $|\sin\phi|$ &    $2/\sin\theta^*$  \\
 \hline
2 &  1 & $  0$ & $|\sin\phi|$ &    $2/\sin\theta^*$  \\
 \hline
3 &  1 & $  0$ & $|\sin\phi|$ &    $2/\sin\theta^*$  \\
 \hline
4 &  1 & $  0$ & $|\sin\phi|$ &    $2/\sin\theta^*$  \\
 \hline
 \end{tabular}
\caption{Leg properties for $T_{m,g}$; $\langle\ln g_\ell(\phi)\rangle$, not
  shown, is $-\ln2$ for all legs.}
\label{tab:Tmg}
}

The main difference between the thrust minor and the transverse thrust
concerns the properties of the outgoing legs, which for $T_{m,g}$ have
$b_\ell=0$ (table~\ref{tab:Tmg}). Referring to
eq.~(\ref{eq:Lg1-expansion}), one sees that this implies larger double
logarithms than for $\tau_{\perp,g}$. Additionally, the dependence on
multiple emissions is more complex, $T_{m,g}$ not having the simple
property of additivity, so it is necessary to calculate $\cF(R')$
numerically. The result, shown in fig.~\ref{fig:tmin-cF}, is similar
(though not identical) for all hard-scattering channels, and uniformly
below $1$, which is indicative of the fact that multiple emissions
increase the value of the observable relative to a single emission.
We also note that the function $\cF$ does not depend on the Born
momenta $\momConf$, as can be verified numerically.

\DOUBLEFIGURE%
{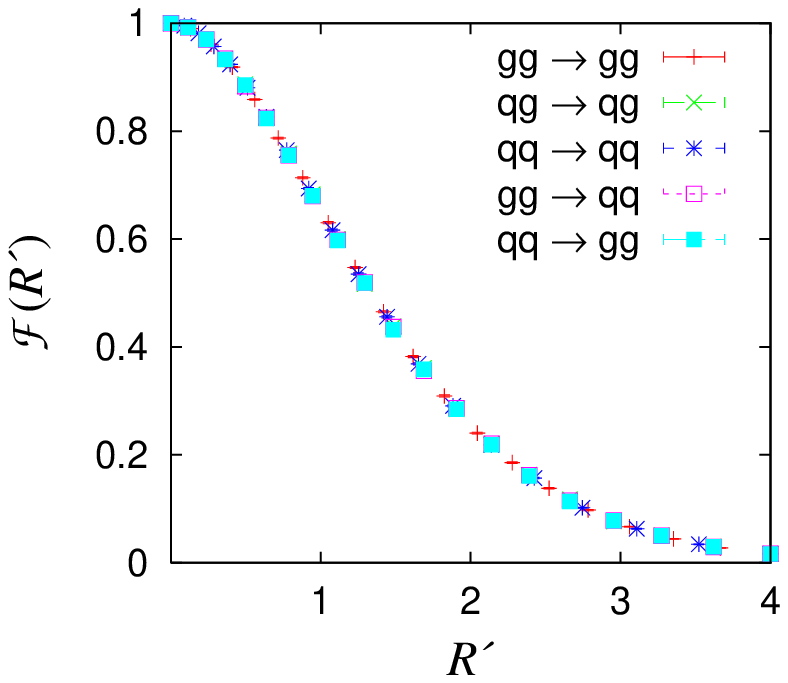,width=0.47\textwidth}%
{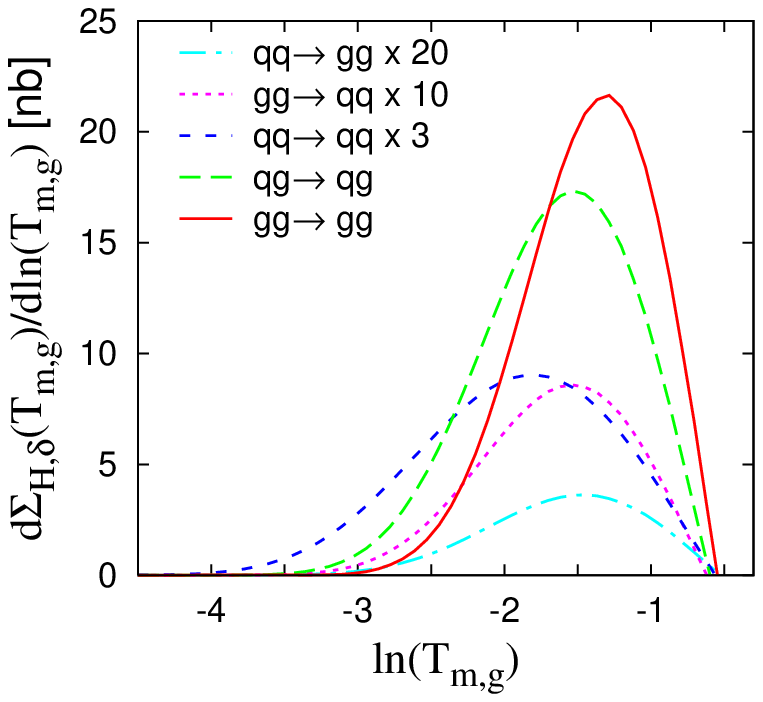,width=0.45\textwidth}%
{The $\cF(R')$ function for $T_{m,g}$.
\label{fig:tmin-cF}}%
{The distribution of $T_{m,g}$ in various hard-scattering channels.
\label{fig:tmin-dists-channels}
}%

The larger double logarithms ($b_{3,4}=0$) and the larger coefficient
$d_\ell$ (together with the larger average value of $g_\ell(\phi)$)
all contribute to $T_{m,g}$ having a distribution,
fig.~\ref{fig:tmin-dists-channels}, that is dominated by
considerably larger values of the observable than was the case for
$\tau_{\perp,g}$. As a result the peak of the distribution is actually at only
\emph{moderately} small values of $T_{m,g}$, where fixed-order
corrections (matching) may have a significant effect on the
distribution.

The rescaling factor, $X$, in the resummation here (of $\ln 1/(X
T_{m,g})$) is $X=1$. As a result the limited detector reach translates
to a limit of applicability of the resummation $\ln T_{m,g} \gtrsim
-\eta_{\max}$. The only channel that extends at all below our
reference value of $\eta_{\max}\simeq 3.5$ is $qq \to qq$,
which however, at these energies does not contribute significantly to
the total distribution.

Finally we observe that the linearity of the definition of $T_{m,g}$
implies that it should have a sensitivity to the underlying event that
is qualitatively quite similar to that of $\tau_{\perp,g}$,
eqs.~(\ref{eq:tautg-underlying-event}).

\subsection{Three-jet resolution threshold}
\label{sec:y3}

Clustering jet algorithms, such as the $k_t$-algorithm
\cite{JetratesHH-CDSW,JetratesHH-EllisSoper,y3-kt_ee}, involve a
successive set of pairwise particle recombinations, so as to group
particles into jets. These algorithms often involve a jet-resolution
parameter, which determines the point at which to stop the pairwise
recombination. The smaller the parameter, the larger the number of
separate jets that gets resolved.

One of the most widely studied observables in $\ee$ is the jet
resolution \emph{threshold}, $y_{23}$, above which the event is
classified as having two jets, and below which it has three (or more)
jets. Though an analogous threshold has been defined for going from
$2+2$ to $2+3$ jet events with the longitudinally invariant
$k_t$-algorithm in hadron-hadron collisions \cite{JetratesHH-CDSW}, it
has received somewhat less attention. Resummed calculations
\cite{JetratesHHResummed} (to lower accuracy than is given here) and
measurements \cite{D0subjets} have been carried out only for the
related problem of subjets within a single jet.

Jet threshold resolutions and jet rates are closely related since the
integrated cross section $\Sigma_\cH(y)$ for $y_{23}$ is equivalent to
the cross section for having a two-jet event, given a jet resolution
of $y$.

We will consider here just one variant of the $k_t$-algorithm (the
exclusive form of that adopted for run II of the Tevatron
\cite{RunII-jet-physics}).  Reference implementations of this and
other schemes are available in \cite{KtJet}.
\begin{enumerate}
\item 
  One defines, for all $n$ final-state (pseudo)particles still in
  the event,  
  \begin{equation}
    d_{kB} = q_{\perp k}^2\,,
  \end{equation}
  and for each pair of final state particles
  \begin{equation}
    \label{eq:dij_hh}
    d_{kl} = \min\{{q_{\perp k}^2,q_{\perp l}^2}\}
    \left((\eta_k-\eta_l)^2+(\phi_k-\phi_l)^2\right)\,.
  \end{equation}
\item One determines the minimum over $k$ and $l$ of the $d_{kl}$ and
  the $d_{kB}$ and calls it $d^{(n)}$. If the smallest value is
  $d_{iB}$ then particle $q_{i}$ is included in the beam and
  eliminated from the final state particles.  If the smallest value is
  $d_{ij}$ then particles $q_{i}$ and $q_{j}$ are recombined into a
  pseudoparticle (jet). A number of recombination procedures exist. We
  adopt the E-scheme, in which the particle four-momenta are simply
  added together,
\begin{equation}
  q_{i j} = q_{i}+q_{j}\,.
\end{equation}
\item The procedure is repeated until only 3 pseudoparticles are left
  in the final state. The observable we resum is then
\begin{equation}
  \label{eq:hhy23}
  y_{23} = \frac{1}{E_{\perp}^2} \max_{n \ge 3}\{d^{(n)}\}\>,
\end{equation}
where $E_{\perp}$ is defined by further clustering the event until
only two jets
remain and taking $E_{\perp}$ as the sum of the two jet transverse
energies,
\begin{equation}
  \label{eq:Eperpdef}
  E_{\perp} = E_{\perp,1} + E_{\perp,2}\,.
\end{equation}
The reason for considering $\max_{n \ge 3}\{d^{(n)}\}$ in
eq.~(\ref{eq:hhy23}), instead of simply $d^{(3)}$, is that the
recombination procedure is not necessarily monotonic in the
$d^{(n)}$ \cite{JetratesHH-CDSW}, so that $n >m$ does not automatically
imply $d^{(n)} < d^{(m)}$.
\end{enumerate}

\TABLE{
 \begin{tabular}{| c | c | c | c | c |}
 \hline
 leg $\ell$ & $a_{\ell}$ & $b_{\ell}$ & $g_{\ell}(\phi)$ & $d_{\ell}$  \\
 \hline
 \hline
1 &  2 & $  0$ & 1 &    $1/\sin^2\theta^*$ \\
 \hline
2 &  2 & $  0$ & 1 &    $1/\sin^2\theta^*$ \\
 \hline
3 &  2 & $  0$ & 1 &    $1/\sin^2\theta^*$ \\
 \hline
4 &  2 & $  0$ & 1 &    $1/\sin^2\theta^*$ \\
 \hline
 \end{tabular}
\caption{Leg properties for $y_{23}$.}
\label{tab:y23}
}

From table~\ref{tab:y23} one sees that, for all legs, the effect of
single emissions scales as the squared transverse momentum, without
any rapidity or azimuthal dependence --- in this respect the
observable is similar to $y_{23}$ in $\ee$.

An interesting theoretical feature of the hadronic $y_{23}$ concerns
$\cF$, fig.~\ref{fig:y23-cF}. In contrast to the transverse thrust and
thrust minor cases, $\cF(R')$ has a strong dependence on the
scattering channel. It seems that along the outgoing legs, multiple
subjets can combine together before being combined with the main hard
jet --- this means that multiple emissions along an outgoing leg lead
to a larger $y_{23}$ than a single emission, causing $\cF$ to be
smaller than $1$, as in $\ee$ \cite{BSZ}. For incoming legs instead,
subjets tend not to recombine together, so that the value of $y_{23}$
is determined exclusively by the hardest subjet. This type of
situation leads to $\cF(R') = 1$. The actual value for $\cF$ then
depends on the relative importance (colour factors) of the incoming
and outgoing jets. 

\DOUBLEFIGURE%
{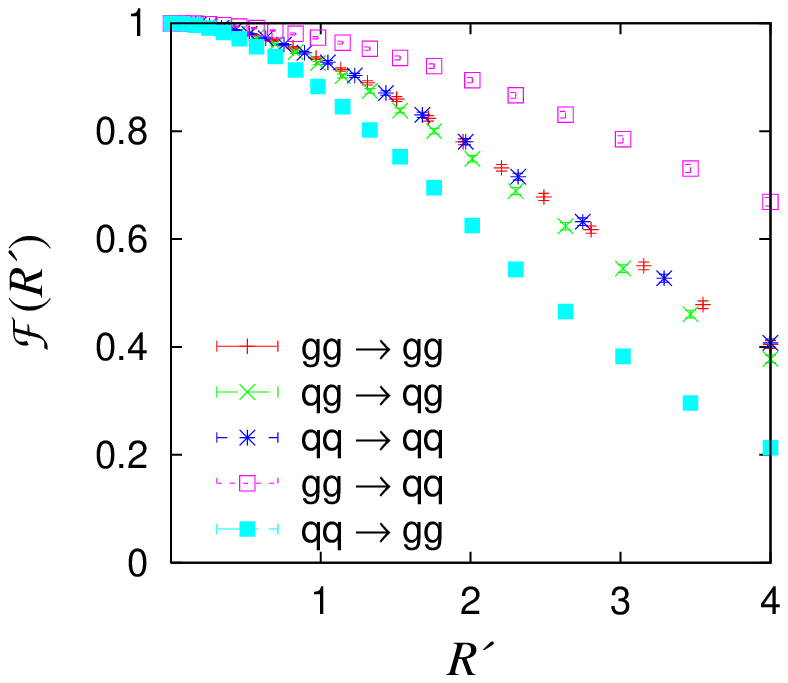,width=0.48\textwidth}%
{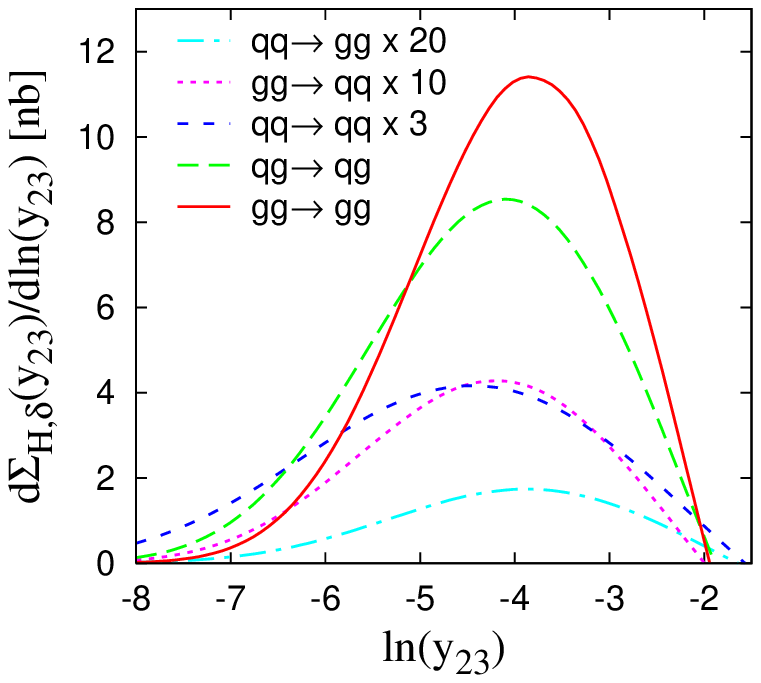,width=0.45\textwidth}%
{The $\cF(R')$ function for $y_{23}$.
\label{fig:y23-cF}}%
{The distribution of $y_{23}$ in various hard-scattering channels.
\label{fig:y23-dists-channels}
}%

Figure~\ref{fig:y23-dists-channels} shows the distribution of $y_{23}$
for different hard subprocesses. It extends to considerably smaller
values than, say, $T_{m,g}$, the reason being that, from the point of
view of LL contributions $y_{23} \sim T_{m,g}^2$. This suggests that
the distribution of $\frac12 \ln y_{23}$ should be comparable to that
of $\ln T_{m,g}$. One sees that to some extent this is the case,
though the $y_{23}$ distribution is somewhat wider than would be
expected based on this argument. This is a consequence of the larger
suppression that $T_{m,g}$ receives from its $\cF(R')$ (note that at a
given value of $y_{23} \sim T_{m,g}^2$, the definition of $R'$ is such
that $R'_{y_{23}} = \half R'_{T_{m,g}}$).

The fact that $y_{23} \sim T_{m,g}^2$ is important also from the point
of view of the impact of the limited experimental rapidity reach,
since the range in which the cut can be ignored theoretically is $\ln
y_{23} \gtrsim -2 \eta_{\max}$. One sees that for $\eta_{\max}=3.5$,
this includes most of the distribution, except in the $qq\to qq$
channel, which extends a little beyond.

While perturbatively $y_{23}$ and $T_{m,g}^2$ are similar, they
should have very different sensitivities to non-perturbative physics.
It is already known that in $\ee$, $k_t$-type jet algorithms are
relatively insensitive to non-perturbative effects. The same should
be true also of the hadronic $y_{23}$ --- for example, the fact that
along incoming legs it is the hardest subjet that dominates, means
that unlike $T_{m,g}$, $y_{23}$ does not `collect' non-perturbative
contributions from the whole range of $|\eta| < \eta_{\max}$, but
rather just from a limited region around the hardest subjet.

\section{Observables with exponentially suppressed forward terms}
\label{sec:rho-brd-etc}

While the observables given above could be easily defined in a
`global' manner, other jet observables, such as jet masses or
broadenings, are more naturally defined just in terms of a central
region, or of the two leading jets. As a result they are usually not
global, which prevents them from being resummed with current automated
approaches such as \caesar.

One solution to this problem is to explicitly add to them a term
sensitive to emissions along the beam direction, so as to render them
global.  If we add just the $q_{\perp i}$, we will obtain observables
whose incoming legs have rather similar properties to the transverse
thrust or the thrust minor, in particular with regards to their
sensitivity to the underlying event and to relevance of the
experimental $\eta_{\max}$ cutoff.

Instead we add a contribution, $\sim q_{\perp i} e^{-|\eta_i|}$, with
an exponential suppression in the forward direction.\footnote{In the
  spirit of \cite{BKS03,BS03} one could also investigate continuous
  classes of observables, in which the forward term goes as $q_{\perp
    i} e^{-c|\eta_i|}$, $c$ being a parameter that allows one to span
  a whole class of
  observables.} %
This will usually
lead to observables having $b_\ell = a$ on the incoming legs, which
doubles the range of $L$, eq.~(\ref{eq:lnV-reach}), in which the
resummed predictions are insensitive to $\eta_{\max}$. Such
observables will usually also have a reduced sensitivity to the
underlying event, since the integral
eq.~(\ref{eq:tautg-underlying-event-int}) will converge rapidly in
$\eta$ giving a result of order $\langle
k_\perp^{(u.e.)}\rangle/(E_{\perp,1}+E_{\perp,2})$, without any
$\eta_{\max}$ enhancement.

Before defining such observables, let us first discuss the central
region, $\cC$, in which the `main' event shape (for example the sum of
jet masses) will be measured.  Various options are possible, for
example taking all particles in the two hardest jets; or taking all
particles having $|\eta_i| < \eta_c + \delta\eta$, where $\eta_c$
specifies the region in which the two hardest jets should lie
(section~\ref{sec:event-select}), while $\delta\eta$ extends this
region, so as to ensure that all selected jets are well contained
within $\cC$. For implementation in \caesar, we have used this second
definition, with $\delta\eta=0.4$, though it should be noted that
since our final variables will be global, the NLL resummed results are actually
independent of the precise definition of $\cC$.\footnote{While at
  higher resummed orders, and in the fixed-order perturbative and the
  non-perturbative contributions there will be dependence on the
  definition of $\cC$.}

Having established $\cC$, we then introduce the mean transverse-energy
weighted rapidity, $\eta_\cC$, of the central region,
\begin{equation}
  \label{eq:etabar}
  \eta_\cC = \frac{1}{
    Q_{\perp,\cC}} \sum_{i\in\cC} \eta_i\, q_{\perp i}\,,\qquad\quad
    Q_{\perp,\cC} = \sum_{i\in \cC} q_{\perp i}\,,
\end{equation}
and define the exponentially suppressed forward term,
\begin{equation}
  \cE_\barcC = \frac{1 }{Q_{\perp,\cC}}
  \sum_{i \notin \cC} q_{\perp i} \,e^{-|\eta_i - \eta_\cC|}\,,
\end{equation}
such that it is invariant with respect to longitudinal boosts (modulo
the dependence of $\cC$ itself on boosts).  We are now in a position
to define some observables.

\subsection{Transverse thrust, minor and jet resolution}
\label{sec:ttmjr-exp}

\paragraph{Transverse thrust.} Let us first define a thrust similar to
that of section~\ref{sec:tau-perp}, but only in terms of the particles
in $\cC$,
\begin{equation}
  \label{eq:Tperp-C}
  T_{\perp,\cC} \equiv \max_{\vec n_{T,\cC}}  \frac{\sum_{i\in\cC} |{\vec
      q}_{\perp i}\cdot {\vec n_{T,\cC}}|}{Q_{\perp,\cC}}\,, 
  \qquad\quad \tau_{\perp,\cC} \equiv 1 - T_{\perp,\cC}\,.
\end{equation}%
\TABLE{
 \begin{tabular}{| c | c | c | c | c |}
 \hline
 leg $\ell$ & $a_{\ell}$ & $b_{\ell}$ & $g_{\ell}(\phi)$ & $d_{\ell}$  \\
 \hline
 \hline
1 & $ 1 $ & $  1$ & 1 & $    1/\sin\theta^* $  \\
 \hline
2 & $ 1 $ & $  1$ & 1 & $    1/\sin\theta ^*$ \\
 \hline
3 & $  1 $ & $  1$ & $\sin^2\phi$ & $  1/\sin^2\theta^* $ \\
 \hline
4 & $  1 $ & $  1$ & $\sin^2\phi$ & $  1/\sin^2\theta^* $ \\
 \hline
 \end{tabular}
\caption{Leg properties for $\tau_{\perp,\cE}$.}
\label{tab:tauperp-E}
}%
We then add the contribution $\cE_\barcC$ with the dependence on the
forward emission,
\begin{equation}
  \label{eq:tauperp-E}
  \tau_{\perp,\cE} \equiv \tau_{\perp,\cC} + \cE_{\barcC}\,,
\end{equation}
to obtain a global observable,\footnote{It actually turns out that
  even without the addition of any direct dependence on particles not
  in $\cC$, $\tau_{\perp,\cC}$ has an indirect sensitivity to them via
  effects of recoil of the hard jets. This leads to $\tau_{\perp,\cC}$
  being discontinuously global ($a_{1,2} = 2$, $a_{3,4}=1$) rather
  than non-global, which is still beyond the scope of \caesar.} %
the resulting leg properties being shown in table~\ref{tab:tauperp-E}.
One sees that for the outgoing legs, they coincide with those of
table~\ref{tab:tautg} for $\tau_{\perp,g}$, while for the incoming
legs $b_\ell=1$, as expected. Like $\tau_{\perp,g}$, the observable is
additive, so $\cF$ is known analytically.

We could at this point continue the discussion for $\tau_{\perp,\cE}$
along lines similar to those for the observables given above. Since,
however, one can imagine introducing quite a few further observables,
we prefer from now on to concentrate on just a subset of them, so as
to illustrate interesting new features. Accordingly we refer the
reader to detailed web pages \cite{qcd-caesar.org} for further
information about $\tau_{\perp,\cE}$, as well as analogous extensions
of the \textbf{thrust minor},
\begin{equation}
  \label{eq:Tm-CE}
  T_{m,\cC} \equiv \frac{1}{Q_{\perp,\cC}} \sum_{i \in \cC}
  |q_{xi}|\,,\qquad\quad
  T_{m,\cE} = T_{m,\cC} + \cE_{\barcC}\,,
\end{equation}
(because of kinematic recoil, it turns out that $T_{m,\cC}$ is global,
and $T_{m,\cE}$ has $b_{1,2}=0$ and identical resummation properties
to $T_{m,\cC}$); and the \textbf{three-jet resolution threshold}, with
$y_{23,\cC}$ defined by the algorithm of section~\ref{sec:y3} applied
only to the final state particles in $\cC$ (and the beam) and
\begin{equation}
  \label{eq:y23-E}
  y_{23,\cE} \equiv y_{23,\cC} + \cE_\barcC^2\,.
\end{equation}
Note that it is necessary to add $\cE_\barcC^2$ here (rather than
$\cE_\barcC$), so as to ensure the continuous globalness of the
observable, $a_1 = a_2 = a_3 = a_4$.

Let us now concentrate in more detail on some observables that are
more naturally defined without explicit reference to a central region.

\subsection{Jet masses}
\label{sec:jet-masses}

Having determined a (central) transverse thrust axis $\vec n_{T,\cC}$
as above, one can separate the central region $\cC$ into an up part
$\cC_U$ consisting of all particles in $\cC$ with $\vec p_{\perp}
\cdot \vec n_{T,\cC} > 0$ and a down part $\cC_D$, particles in $\cC$ with
$\vec p_{\perp} \cdot \vec n_{T,\cC} < 0$. If $\cC$ is taken to be made up of
all particles in the two hardest jets, one can also define $\cC_U$ and
$\cC_D$ as consisting of the two jets separately.  Such an alternative
definition should not change any of the NLL resummed predictions for
the global variables, and may actually help to minimise subleading (NNLL)
corrections.\footnote{In contrast, for the central, non-global
observables (those with a $\cC$ suffix), the exact choice of $\cC$
will affect the NLL terms.} %

One then defines, in analogy with $\ee$ \cite{Clavelli}, the
normalised squared invariant masses of the two regions\footnote{For
  certain non-perturbative studies, it can be advantageous
  \cite{SalamWicke} to use a so-called `$E$-scheme' definition in
  which the $3$-momenta $\vec q_i$ are rescaled $\vec q_i \to
  (E_i/|\vec q_i|) \vec q_i$, equivalent to a massless approximation
  for all particles. }
\begin{equation}
  \label{eq:mass-XC}
  \rho_{X,\cC} \equiv \frac{1}{Q_{\perp,\cC}^2}
  \left(\sum_{i\in \cC_X} q_{i}\right)^2\,,\qquad X = U, D\,,
\end{equation}
from which one can obtain a (non-global) central sum of masses and
heavy-mass,
\begin{equation}
  \label{eq:mass-C-sum-heavy}
  \rho_{S,\cC} \equiv \rho_{U,\cC} + \rho_{D,\cC}\,,\qquad\quad
  \rho_{H,\cC} \equiv \max\{\rho_{U,\cC}, \rho_{D,\cC}\}\,,
\end{equation}
together with versions that include the addition of the
exponentially-suppressed forward term,
\begin{equation}
  \label{eq:mass-E-sum-heavy}
  \rho_{S,\cE} \equiv \rho_{S,\cC} + \cE_{\barcC}\,,\qquad\quad
  \rho_{H,\cE} \equiv \rho_{H,\cC} + \cE_{\barcC}\,.
\end{equation}
\TABLE{
 \begin{tabular}{| c | c | c | c | c |}
 \hline
 leg $\ell$ & $a_{\ell}$ & $b_{\ell}$ & $g_{\ell}(\phi)$ & $d_{\ell}$  \\
 \hline
 \hline
1 & $ 1 $ & $  1$ & 1 & $    1/\sin\theta^* $ \\
 \hline
2 & $ 1 $ & $  1$ & 1 & $    1/\sin\theta^* $ \\
 \hline
3 & $  1 $ & $  1$ & 1 & $    1/\sin^2\theta^* $ \\
 \hline
4 & $  1 $ & $  1$ & 1 & $    1/\sin^2\theta^* $ \\
 \hline
 \end{tabular}
\caption{Leg properties for $\rho_{S,\cE}$ (and $\rho_{H,\cE}$).}
\label{tab:rhosum-E}
}%
The single-emission leg properties for $\rho_{S,\cE}$ and
$\rho_{H,\cE}$ are identical (even beyond the soft-collinear limit),
and quite similar to those of $\tau_{\perp,\cE}$ except for the lack
of azimuthal dependence on the outgoing legs. The sum of squared
masses, $\rho_{S,\cE}$, is an additive observable, so $\cF(R')$ is
given by eq.~(\ref{eq:additive-cF}), while for the heavy-mass it needs
to be computed numerically (and decreases for increasing $R'$).

\EPSFIGURE{
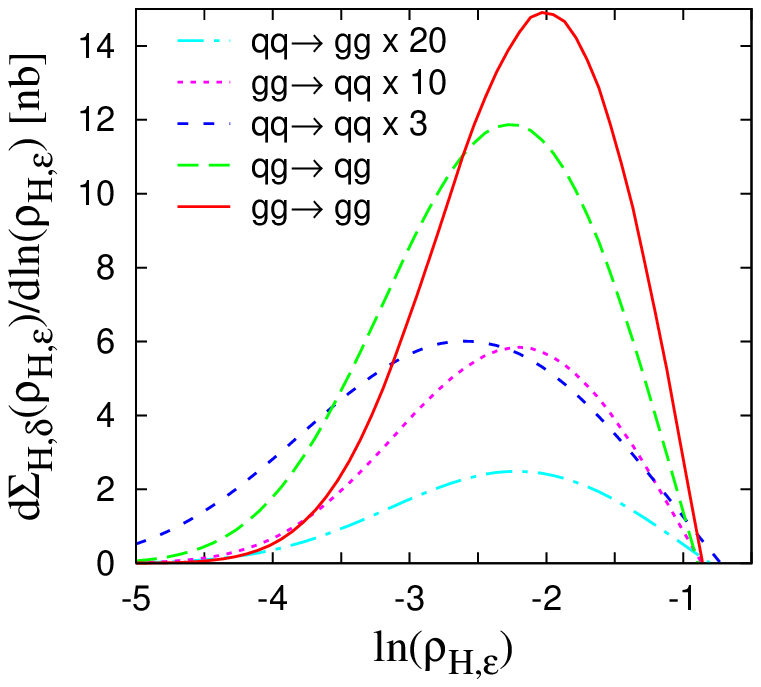,
width=0.45\textwidth
}{The distribution of $\rho_{H,\cE}$ in the different hard-scattering
  channels.\label{fig:rhoHE-dists-channels}}
The resummed distribution for $\rho_{H,\cE}$ is shown,
channel-by-channel in fig.~\ref{fig:rhoHE-dists-channels}. Compared,
say, to an observable like $T_{m,g}$, the distributions extend to
relatively low values, a consequence of the smaller LL terms (since
$b=a$). However, as mentioned at the beginning of this section, the
fact that $b_{1,2}=a=1$ means that the limit of applicability of the
resummation, in the presence of an experimental rapidity cut, is given
by $\ln \rho_{H,\cE} \gtrsim -2\eta_{\max}$. Thus all channels for the
distribution are comfortably contained within this region.

\subsection{Jet Broadenings}
\label{sec:broadenings}

With the same division into up and down regions as for the jet masses,
one can define jet broadenings. To do so in a boost-invariant manner,
one first introduces rapidities and azimuthal angles of axes for the
up and down regions,
\begin{equation}
  \label{eq:broadening-axes}
  \eta_{X,\cC} \equiv \frac{\sum_{i\in \cC_X} q_{\perp i}
    \eta_i}{\sum_{i\in \cC_X} 
    q_{\perp i}}\,,\qquad
  \phi_{X,\cC} \equiv \frac{\sum_{i\in \cC_X} q_{\perp i}
    \phi_i}{\sum_{i\in \cC_X} 
    q_{\perp i}}\,,\qquad
   X = U, D\,,
\end{equation}
and defines broadenings for the two regions,
\begin{equation}
  \label{eq:BX-C}
  B_{X,\cC} \equiv \frac{1}{2Q_{\perp,\cC}} \sum_{i\in\cC_X}
  q_{\perp i}\sqrt{(\eta_i-\eta_{X,\cC})^2 + (\phi_i - \phi_{X,\cC})^2}\,,
  \qquad X = U, D\,,
\end{equation}
from which one can obtain central total and wide-jet broadenings,
\begin{equation}
  \label{eq:B-C-total-wide}
  B_{T,\cC} \equiv B_{U,\cC} + B_{D,\cC}\,,\qquad\quad
  B_{W,\cC} \equiv \max\{B_{U,\cC}, B_{D,\cC}\}\,.
\end{equation}
Adding the exponentially-suppressed forward terms gives global
observables, 
\begin{equation}
  \label{eq:BTWE}
  B_{T,\cE} \equiv B_{T,\cC} + \cE_{\barcC}\,,\qquad
  B_{W,\cE} \equiv B_{W,\cC} + \cE_{\barcC}\,.
\end{equation}
\TABLE{
 \begin{tabular}{| c | c | c | c | c |}
 \hline
 leg $\ell$ & $a_{\ell}$ & $b_{\ell}$ & $g_{\ell}(\phi)$ & $d_{\ell}$  \\
 \hline
 \hline
1 & $ 1 $ & $  1$ & 1 & $    1/\sin\theta^* $  \\
 \hline
2 & $ 1 $ & $  1$ & 1 & $    1/\sin\theta^* $ \\
 \hline
3 & $  1 $ & $  0$ & 1 & $    1/\sin\theta^* $ \\
 \hline
4 & $  1 $ & $  0$ & 1 & $    1/\sin\theta^* $ \\
 \hline
 \end{tabular}
\caption{Leg properties for $B_{T,\cE}$ (and $B_{W,\cE}$).}
\label{tab:BTW-E}
}%
The single-emission properties of these two observables are identical
and are shown in table~\ref{tab:BTW-E}. In both cases $\cF(R')$ needs
to be calculated numerically, and it decreases with increasing $R'$
(more strongly for $B_{T,\cE}$).

The distribution for $B_{W,\cE}$, separated into channels, is shown in
fig.~\ref{fig:BWE-dists-channels}. As expected from the different
double logarithmic structure for the outgoing legs, the $B_{W,\cE}$
distributed is centred at larger values of the observable than
$\rho_{H,\cE}$. This can be seen clearly also from
fig.~\ref{fig:MH-BW-dists-sum} which shows both distributions, summed
over channels for two different $E_{\perp,\min}$ cuts. Since a limited
experimental rapidity translates into an allowed range of the
observable's value $\ln
v \gtrsim -2\eta_{\max}$, we see that the distributions for both
$\rho_{H,\cE}$ and $B_{W,\cE}$ are within the theoretically reliable
region. The same statement applies to most of the other observables
defined in this section.

\DOUBLEFIGURE%
{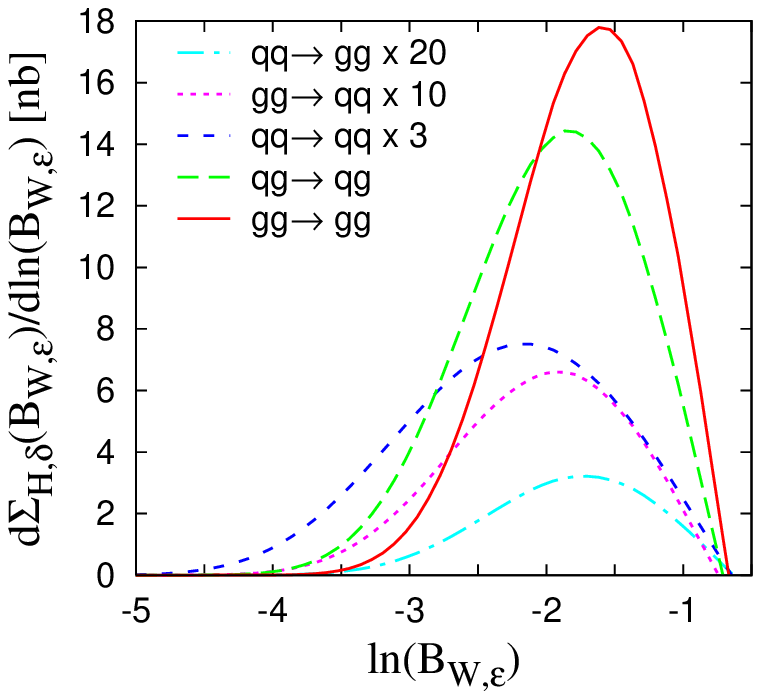,width=0.48\textwidth
}%
{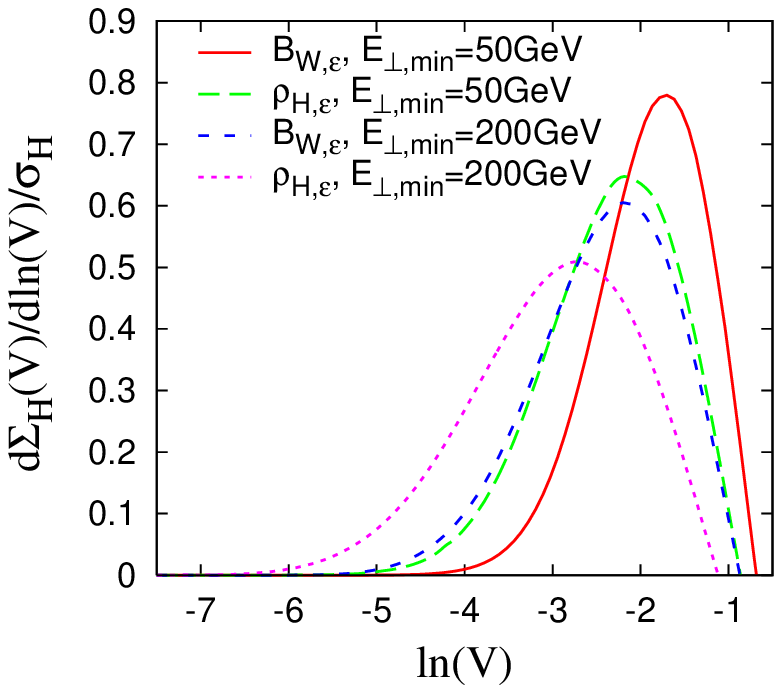,width=0.48\textwidth
}%
{The distribution of $B_{W,\cE}$ for separate hard-scattering channels.
\label{fig:BWE-dists-channels}}%
{The distributions for $\rho_{H,\cE}$ and $B_{W,\cE}$, summed over
  channels, for two $E_{\perp,\min}$ cuts.
\label{fig:MH-BW-dists-sum}}%

\section{Indirectly global observables}
\label{sec:ind-glob}

We call `indirectly global' those observables that explicitly measure
only a subset of the particles in an event, but are nevertheless
indirectly sensitive to the remaining emissions, typically through
recoil.  Previously studied examples include certain DIS Breit-frame
event shapes, such as the current-hemisphere broadening with respect
to the photon axis~\cite{DSBroad} or one of the variants of the
thrust, $\tau_{zQ}$~\cite{ADS}.

Indirectly global observables have the advantage that they can be
defined purely in terms of particles in the central region $\cC$,
while still remaining global. This eliminates potential problems
associated with the limited experimental reach in rapidity.

To construct an indirectly global observable, one takes one of the
`central', non-global observables of the previous section, and then
adds to it some second `recoil' quantity, also defined only in terms
of the momenta of $\cC$, but that is sensitive to the emissions outside
$\cC$. An example of such a `recoil' quantity is the 2-dimensional
vector sum of the transverse momenta in $\cC$,
\begin{equation}
  \label{eq:recoil-term}
  \cR_{\perp,\cC} \equiv \frac{1}{Q_{\perp,\cC}} \left|\sum_{i \in \cC} {\vec
        q}_{\perp i}\right|\,,
\end{equation}
(with $Q_{\perp,\cC}$ defined in eq.~(\ref{eq:etabar})) which, by conservation of momentum, is equal to (minus) the vector sum of the transverse momenta outside $\cC$. Thus we obtain a whole series of `indirectly global' event shapes, whose definitions are similar to those of section~\ref{sec:rho-brd-etc}, but with $\cE_{\barcC}$ replaced by $\cR_{\perp,\cC}$:
\begin{subequations}
\begin{align}
  \label{eq:tauperp-R}
  \tau_{\perp,\cR} &\,\equiv\, \tau_{\perp,\cC} + \cR_{\perp,\cC}\,,
  \\
  \label{eq:Tm-R}
  T_{m,\cR} &\,\equiv\, T_{m,\cC} + \cR_{\perp,\cC}\,,
  \\
  \label{eq:y23-R}
  y_{23,\cR} &\,\equiv\, y_{23,\cC} + \cR_{\perp,\cC}^2\,,
  \\
  \label{eq:mass-R-sum-heavy}
  \rho_{S,\cR} &\,\equiv\, \rho_{S,\cC} + \cR_{\perp,\cC}\,,\;\,\qquad\quad
  \rho_{H,\cR} \,\equiv\, \rho_{H,\cC} + \cR_{\perp,\cC}\,,
  \\
  \label{eq:BTWR}
  B_{T,\cR} &\,\equiv\, B_{T,\cC} + \cR_{\perp,\cC}\,,\qquad\quad
  B_{W,\cR} \,\equiv\, B_{W,\cC} + \cR_{\perp,\cC}\,.
\end{align}
\end{subequations}

Experimentally, we envisage that the main difficulty that will arise
specifically for this class of observables is the accurate
determination of $\cR_{\perp,\cC}$, since it involves a cancellation
between hard momenta. The extent to which this `missing transverse
momentum' can be well measured will determine the extent to which it
will be possible to study the region of low event-shape values.

\DOUBLETABLE{
 \begin{tabular}{| c | c | c | c | c | c |}
 \hline
 leg $\ell$ & $a_{\ell}$ & $b_{\ell}$ & $g_{\ell}(\phi)$ & $d_{\ell}$ \\
 \hline
 \hline
1 & $ 1 $ & $  0$ & 1 & $    1/\sin\theta^* $ \\
 \hline
2 & $ 1 $ & $  0$ & 1 & $    1/\sin\theta^* $ \\
 \hline
3 & $  1 $ & $  1$ & $\sin^2\phi$ & $    1/\sin^2\theta^* $ \\
 \hline
4 & $  1 $ & $  1$ & $\sin^2\phi$ & $    1/\sin^2\theta^* $ \\
 \hline
 \end{tabular}
}%
{
 \begin{tabular}{| c | c | c | c | c | c |}
 \hline
 leg $\ell$ & $a_{\ell}$ & $b_{\ell}$ & $g_{\ell}(\phi)$ & $d_{\ell}$ \\
 \hline
 \hline
1 & $  1 $ & $  0$ & $(1\!+\!|\sin\phi|)/2^*$ & $  2/\sin\theta^* $  \\
 \hline
2 & $  1 $ & $  0$ & $(1\!+\!|\sin\phi|)/2^*$ & $  2/\sin\theta^* $  \\
 \hline
3 & $  1 $ & $  0$ & $|\sin\phi|$ & $    2/\sin\theta^* $ \\
 \hline
4 & $  1 $ & $  0$ & $|\sin\phi|$ & $    2/\sin\theta^* $ \\
 \hline
 \end{tabular}
}%
{Leg properties for $\tau_{\perp,\cR}$.\label{tab:tauperp-R}}%
{Leg properties for $T_{m,\cR}$; $\langle \ln g_{1,2}(\phi)\rangle=
  4G/\pi-2\ln2$. \label{tab:Tm-R}}

The single-emission properties for two of these observables are
illustrated in tables~\ref{tab:tauperp-R} and \ref{tab:Tm-R}. As
expected, the outgoing legs have identical properties to the global
variants of the observables. The incoming legs have $b_\ell=a$, and a
simple calculation reveals that $d_\ell=1/\sin\theta$ and
$g_\ell(\phi)=1$ for most observables, including $\tau_{\perp,\cR}$.
An exception is $T_{m,\cR}$, for which there is an interplay between
recoil dependence present (implicitly) in $T_{m,\cC}$, and the
additional $\cR_{\perp,\cC}$ term.

\DOUBLEFIGURE%
{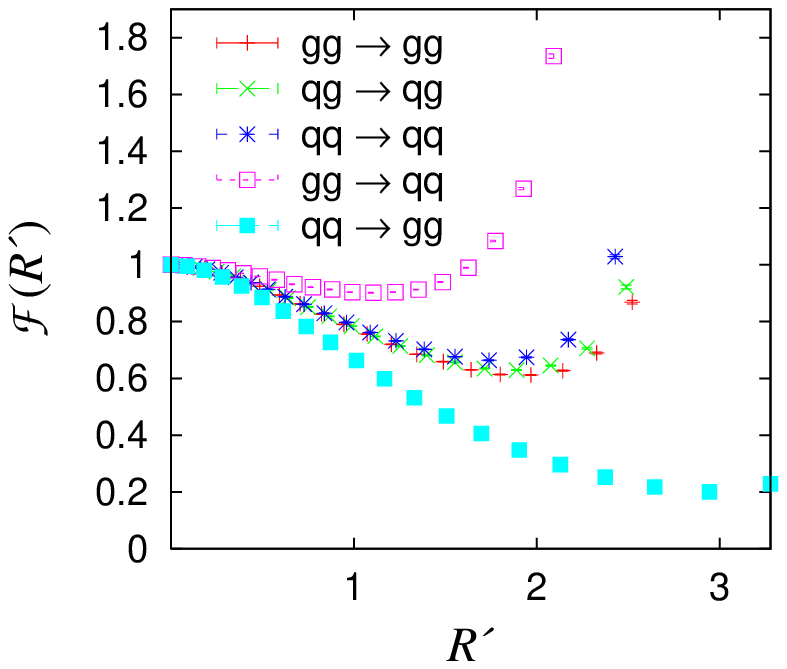,width=0.48\textwidth
}%
{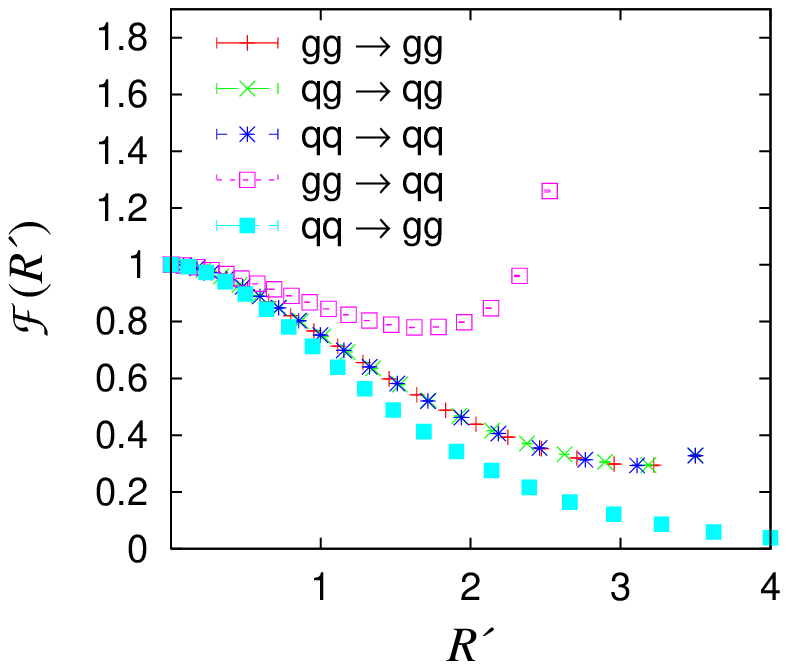,width=0.48\textwidth
}%
{$\cF(R')$ for $\tau_{\perp,\cR}$.
\label{fig:tauperp-R-cF}}%
{$\cF(R')$ for $T_{m,\cR}$.
\label{fig:Tm-R-cF}}%

The $\cF(R')$ functions for $\tau_{\perp,\cR}$ and $T_{m,\cR}$ are
shown in figures \ref{fig:tauperp-R-cF} and \ref{fig:Tm-R-cF}. For
small values of $R'$, $\cF(R')$ is below $1$, as for the observables
discussed in the previous sections. However for sufficiently large
$R'$, $\cF(R')$ starts to increase and it has a divergence at some
finite value of $R' = R'_c$. Such divergences are
actually a well-understood phenomenon \cite{DSBroad,RakowWebber},
characteristic of observables for which contributions from multiple
emissions can cancel. Schematically, they occur because for
sufficiently large $R'$ it becomes more favourable to suppress the
value of the observable via cancellations between emissions than by
Sudakov suppression, and the resulting change in parametric behaviour
of $\Sigma_\cH(v)$ cannot be represented as a subleading correction to
a Sudakov resummation.

For the observables being discussed here, the cancellation is in the
vector sum of transverse momenta.  This being a two-dimensional
cancellation, the divergence is located at
$R'_{1+2}=2$~\cite{Caesar,DSBroad}, where $R'_{1+2}$ is the part of
$R'$ associated with the incoming legs. For cases where $R'_{1+2} =
\frac12 R'$, such as the $gg\to gg$ channel for $T_{m,\cR}$ then
$R'_c=4$. For the $gg\to qq$ scattering channel, in which $R'$ comes
mostly from the incoming legs, the divergence is at lower $R'$.
Furthermore for any given fixed channel, the divergence occurs earlier for
the transverse thrust than for the minor, because the thrust has
smaller LL terms, and accordingly a larger proportion of $R'$ is
associated with the incoming legs.

The divergence is a structure that appears even more strongly at yet
higher orders (NNLL, \ldots) and it could in principle be resummed.
Technically speaking, one would have to carry out a $b$-space
resummation \cite{b-space} on the incoming legs, a `normal'
resummation for the outgoing legs, and include a non-global
resummation \cite{NG1} to deal with the dynamic discontinuous
non-globalness \cite{DiscontGlobal} associated with the boundary of
$\cC$. This is technically rather challenging and so far the only such
resummation \cite{DSBroad}, for a jet-broadening in DIS, neglected the
non-global logarithms. The conclusion of that study (for a case in
which $R'_c=4$) was that the further resummation had negligible
practical impact except in a region where the distribution was already
very strongly suppressed.

\DOUBLEFIGURE%
{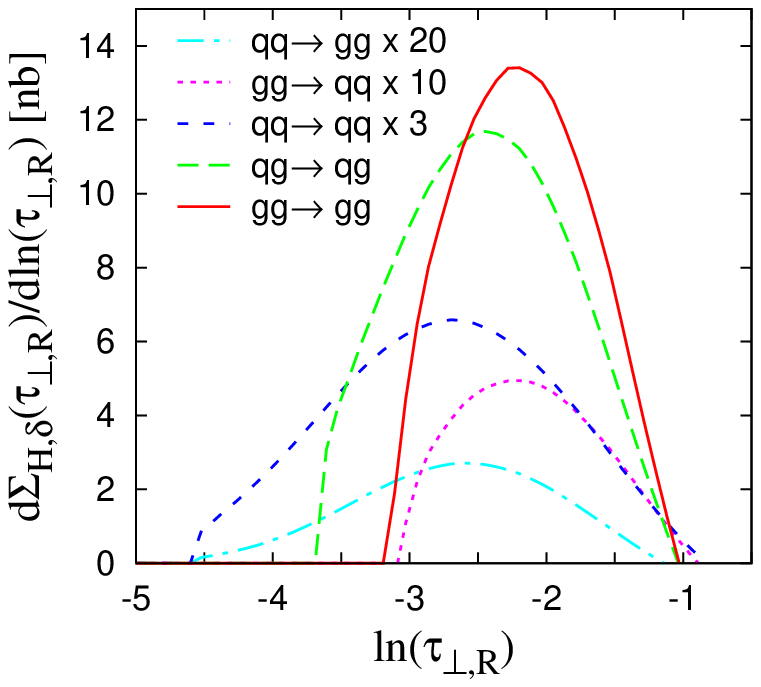,width=0.48\textwidth
}%
{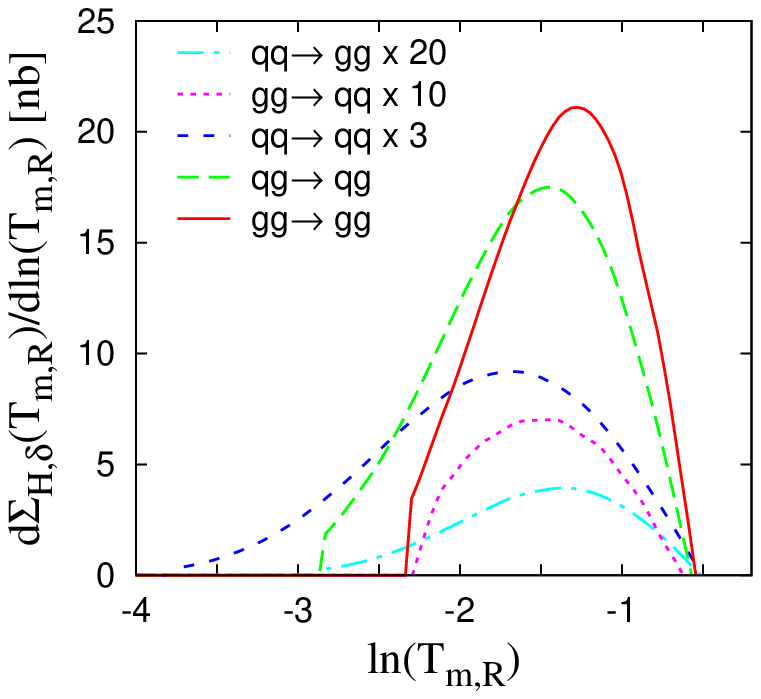,width=0.48\textwidth
}%
{The distribution of $\tau_{\perp,\cR}$ for separate hard-scattering
  channels.
\label{fig:tauperp-R-dists-channels}}%
{The distribution of $T_{m,\cR}$ for separate hard-scattering
  channels.
\label{fig:Tm-R-dists-channels}}%

We therefore follow the approach of \cite{DSBroad} and consider the
distributions for the observables without any additional resummation,
figs.~\ref{fig:tauperp-R-dists-channels} and
\ref{fig:Tm-R-dists-channels}. The distributions are cut (somewhat
arbitrarily) at the point where $R' = 7/8 R'_c$ --- one sees that this
corresponds to different values of $\ln v$, depending on the
resummation channel. However for each channel most of the cross
section is in the theoretically reliable region where the divergence
can be ignored. 
We note that some problems do arise in the combination of channels
with divergences at different points, notably from the point of view
of establishing the region in which the sum over all channels is well
estimated. The detailed discussion of this issue will be left to
future work \cite{BSZMatching}.

We conclude this section by observing that there are reasons to
believe that these observables, though global, should have a relatively
limited sensitivity to the underlying event.  This is because in the
sum of perturbative and `underlying-event' contributions to
$\cR_{\perp,\cC}$, after integration over azimuthal angles, the leading
piece of the underlying event component averages to zero, leaving only
a subleading component.\footnote{Similar arguments have been applied
  for non-perturbative studies of jet-broadenings in $\ee$ and DIS
  \cite{DSBroad,mil2}, with a good degree of phenomenological success
  \cite{DiscontGlobal,BroadNP-rev,MovillaFernandez:2002hu,Kluge:2003sa}.
  The argument assumes a lack of azimuthal correlation between the
  underlying event and perturbative radiation from the `hard' event. }

\section{Conclusions}

In this article we have defined and resummed a set of dijet
event-shape observables and jet rates that are suitable for study at
hadronic colliders, both from a theoretical and an experimental point
of view.  The principle issues that had to be reconciled were those of
globalness, which greatly simplifies the resummed theoretical
predictions, and of limited experimental reach in rapidity. At first
sight these requirements seem contradictory, however we saw that it is
possible to define several classes of observables which meet both
criteria. 

For a majority of the observables, the distribution is
concentrated in a region where the limited experimental rapidity reach
can be ignored in the resummed prediction. This is the case for nearly
all the directly global observables of section \ref{sec:dir-glob} and
those with exponentially suppressed forward terms,
section~\ref{sec:rho-brd-etc}, the latter having been specifically
designed to be optimal in this respect. The observables of
section~\ref{sec:ind-glob}, defined only in terms of particles in the
central region, avoid the problem altogether, implementing globalness
`indirectly' via a recoil term (though this affects the validity of
the resummation in a limited region of very small event-shape values).

The study of a range of observables, as presented here, would have been
considerably more difficult without the help of the automated
resummation tool, \caesar. For brevity we omitted a range of further
results \cite{qcd-caesar.org}. Full phenomenological studies will
necessitate also matching with fixed order calculations such as
\cite{NLOJET,Nagy03,TriRad}, some relevant issues for `practitioners'
having been discussed in appendix~\ref{sec:fixed-order}.

One of the main reasons for examining several classes of observable
(and a number of observables within each class) is that they have
quite complementary sensitivities to different physics issues, both
perturbative and non-perturbative.  Possible studies include
measurements of $\as$, tests of novel perturbative QCD colour
evolution structures that arise in events with $4$-jet topology, and
investigations into hadronisation effects and the properties of the
underlying event. Overall, we believe that this represents a
potentially far richer programme of investigations than has, for
example, been carried out in $\ee$ or DIS processes.


\paragraph{Acknowledgements}

This paper has evolved from discussions with a number of colleagues,
including in particular Leonard Christofek, Yuri Dokshitzer, Joey
Huston, Pino Marchesini, Aurore Savoy-Navarro and Mike Seymour. We are
grateful also the CERN and the IPPP, Durham for the use computing
facilities. GPS would like to thank also the Fermilab theory group for
hospitality while this article was being prepared.

\appendix

\section{Scale choices}
\label{sec:scale-choices}

As discussed in \cite{DiscontGlobal,DSBroad,Jones:2003yv}, in a
resummation at NLL accuracy there is an intrinsic ambiguity in the
choice of the logarithm to be resummed, for instance instead of
resumming
\begin{equation}
\label{eq:lnv0}
        L= \ln\frac{1}{v}\,,
\end{equation}
one could equally well want to resum
\begin{equation}
\label{eq:lnvmod}
        \bar L= \ln\frac{1}{X v}\,,\qquad \mbox{with}\quad X= \cO{1}\,. 
\end{equation}
This alternative choice for the logarithm affects the functional form
of the single-logarithmic terms, but not the overall answer at NLL.
Changing $X$ allows one then to estimate the size of a specific class
of higher order (NNLL) corrections.  In two-jet event shapes, the
default value for $X$ was chosen
in a `natural' way by requiring that the coefficient $G_{11}$, the
$\cO{\as}$ single logarithmic term,
contained only terms due to hard collinear splitting. This
corresponded to taking $X = 1/d$ (usually $d_1 = d_2 \equiv d$ in
$\ee$) and ensured that one didn't introduce large spurious subleading
logarithms whose coefficients involved powers of $\ln d$, associated
just with the overall (arbitrary) normalisation of the observable.

The equivalent procedure more generally would be
to define $X$ so as to cancel the coefficient $d_\ell g_\ell(\phi)$
in eq.~(\ref{eq:parametricform}). The obvious difficulty though is
that $d_\ell g_\ell(\phi)$ is not a constant --- it depends on the
hard leg $\ell$ and on the value of the azimuthal angle.

Inspecting the detailed resummation formulae, in particular eq.~(3.6)
of \cite{Caesar}, one sees that the problem of the azimuthal
dependence can be eliminated, insofar as the resummed result depends
only on $\ln \bar d_\ell \equiv \ln d_\ell + \int\frac{d\phi}{2\pi} \ln
g_\ell(\phi)$. Furthermore, for each leg (if all the $b_\ell$ values
are the same), the term involving $\ln \bar d_\ell$ comes in with a
weight proportional to the colour factor, $C_\ell$, of the leg. So for
$X$ to cancel the coefficient $d_\ell g_\ell(\phi)$ on average, it
suffices to take
\begin{equation}
\label{eq:Xproddparcwt}
        \ln X = -\frac{1}{C_T}\sum_{\ell=1}^{n} \left( C_\ell \ln \bar
        d_\ell\right)\,,\qquad\quad C_T \equiv \sum_{\ell=1}^n C_\ell\,.
\end{equation}
with $n$ the number of hard legs. This choice means however that $L$
is defined differently for different colour configurations, which
while legitimate, may seem a little unnatural.  In this study we
therefore adopt a prescription for $X$, which is numerically almost
indistinguishable from the choice in eq.~(\ref{eq:Xproddparcwt}), but
does not have the unusual feature that the logarithm depends on the
specific colour configuration,
\begin{equation}
\label{eq:Xproddpar}
        \ln X = -\frac{1}{n}\sum_{\ell=1}^{n}\ln \bar d_\ell\,. 
\end{equation}
For situations in which all legs have the same colour factor, it is
equivalent to eq.~(\ref{eq:Xproddparcwt}).

\section{Comparison to fixed order}
\label{sec:fixed-order}

A valuable cross-check on any resummed prediction is that its
expansion to fixed order coincides with the logarithmically enhanced
terms of exact fixed-order predictions.  For $\ee\to2\jets$ and DIS
$1+1\jet$ resummations, such checks have been carried out to NLO
accuracy. The current status of the NLO codes suitable for NLO
prediction of hadronic dijet event shapes is that NLOJET++
\cite{NLOJET} is available publicly, while TRIRAD \cite{TriRad} can be
requested from the authors.

We have carried out comparisons with NLOJET++.\footnote{We are
  grateful to Zoltan Nagy for having provided us with a prerelease of
  a new version of his code.} This turned out to be feasible at LO
(for the $\as L^2$ and $\as L$ terms of $\Sigma_\cH(v)/\sigma_\cH$),
while we encountered difficulties in obtaining a determination of the
NLO contribution that was sufficiently accurate to enable a meaningful
comparison with the expansion of the resummations.  Nevertheless even
the LO comparisons represent a non-trivial check of the resummations,
in particular in view of the involved four-jet large-angle colour
structure.

Fixed-order calculations are relevant not only as checks of the
resummation, but more importantly for matching~\cite{CTTW}, so as to
obtain predictions that are valid for large as well as small values of
the observable. The matching also gives a partial improvement in the
accuracy at small values of the observable, supplementing, for instance,
eq.~(\ref{eq:vProb-general}) with a term known as $C_1$, which depends
on the momentum configuration and the channel,
\begin{equation}
  \label{eq:vProb-general-C1}
  \vProb_{\momConf,\subProc}(v) = \left(1 + C_{1,\momConf,\subProc}
    \frac{\as}{2\pi} +  
  \ldots \right) \exp\left[ L
  g_{1,\subProc}(\as L) + g_{2,\momConf,\subProc}(\as L) 
  + \as g_{3,\momConf,\subProc}(\as L) + \cdots\right]\,,
\end{equation}
where we have explicitly indicated the dependence of the $g_n(\as L)$
on the channel $\subProc$ and (for $n\ge 2$) on the momentum
configuration $\momConf$ (elsewhere this dependence has been left
implicit).

The improvement in accuracy comes, for example, because the product of
$C_{1,\momConf,\subProc}$ and the first term in the expansion of
$Lg_{1,\subProc}(\as L)$ helps to fix the $\as^n L^{2n-2}$ terms in the
expansion of $\vProb_{\momConf,\subProc}(v)$, or equivalently, after
summing over channels and integrating over Born configurations, in
$\Sigma_\cH(v)/\sigma_\cH$.  Currently however, fixed-order programs
are at best able to provide a weighted sum of
$C_{1,\momConf,\subProc}$ values across all channels
\begin{equation}
  \label{eq:sum-C1}
   \langle C_{1,\momConf} \rangle_\delta = \frac{\sum_{\subProc}
     \frac{d\sigma_\subProc}{d\momConf} C_{1,\momConf,\subProc}}{
     \sum_{\subProc}
     \frac{d\sigma_\subProc}{d\momConf}   }\,.
\end{equation}
Since different channels have different colour factors appearing in $L
g_{1,\subProc}(\as L)$, the averaging over channels for
$C_{1,\momConf,\subProc}$ means that one cannot reconstruct the full
information for the sum of products of $C_{1,\momConf,\subProc}$ and
$L g_{1,\subProc}(\as L)$
\begin{equation}
  \label{eq:bad-av-C1-G12}
  \langle\, C_{1,\momConf}\; L g_1(\as L) \,\rangle_\delta \,\ne \,
  \langle C_{1,\momConf}  \rangle_\delta\;
  \langle L g_1(\as L) \rangle_\delta\,.
\end{equation}
The fact that the $C_{1,\momConf,\subProc}$ are characteristics of the
soft and collinear limit, means that they can in principle be
unambiguously extracted channel-by-channel \cite{BSZMatching},
enabling a proper average to be carried out.  However the identification
of the channel requires that one have information on the flavour of
individual partons in the fixed-order calculation. Though such
information is present in some form in existing fixed-order codes
\cite{NLOJET,TriRad}, it is not, as far as we understand, available
through the external `user' interfaces to those codes. The
availability of a method to access this information more
straightforwardly would be of considerable help for the matching.

\end{document}